\documentclass[journal]{vgtc}                 

\graphicspath{{figures/}}

\usepackage{times}

\usepackage{mathptmx}
\usepackage{amsmath}
\usepackage{bm}
\usepackage{booktabs}
\usepackage{threeparttable}
\usepackage{subcaption}
\usepackage{soul}
\usepackage{multirow}

\onlineid{1328}

\vgtccategory{Perception and Cognition}

\ieeedoi{10.1109/TVCG.2025.3549570}

\title{Brain Signatures of Time Perception in Virtual Reality}

\author{
  \authororcid{Sahar Niknam}{0000-0002-8021-1021},
  \authororcid{Saravanakumar Duraisamy}{0000-0002-8691-4991},
  \authororcid{Jean Botev}{0000-0001-8492-7708}, and 
  \authororcid{Luis A. Leiva}{0000-0002-5011-1847}
}

\authorfooter{
  \item
  	Sahar Niknam, Saravanakumar Duraisamy, Jean Botev, and Luis A. Leiva are with the University of Luxembourg.
  	E-mail: \{sahar.niknam\,$|$\,saravanakumar.duraisamy\,$|$\,jean.botev\,$|$\,luis.leiva\}@uni.lu\,.
}

\teaser{
  \centering
  \hspace{-4mm}
  \begin{subfigure}[t]{0.335\linewidth}
    \centering
    \includegraphics[height=14.7em]{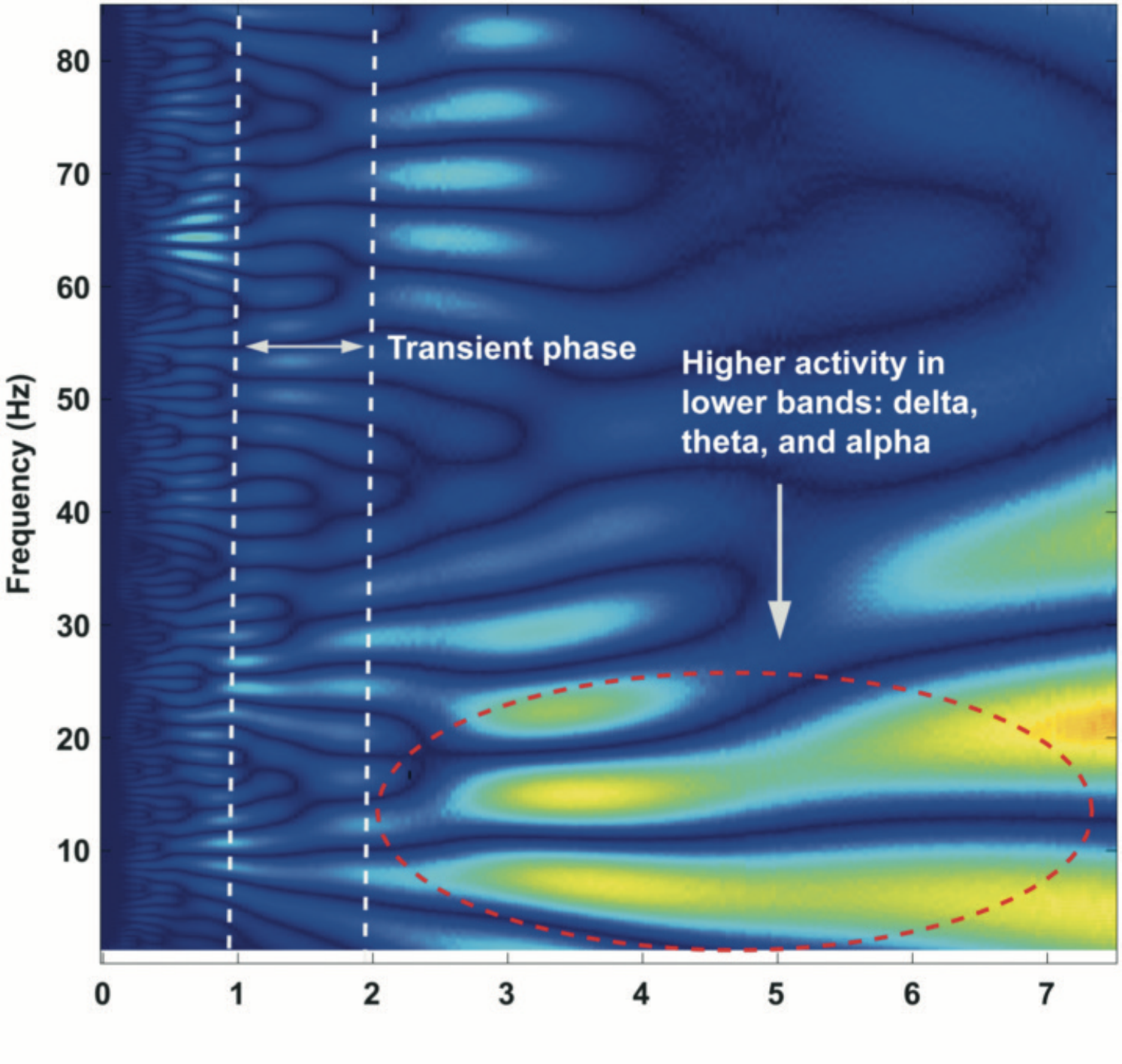}
    \caption{Overestimation.}
    \label{fig:overestimation}
  \end{subfigure}
  \begin{subfigure}[t]{0.325\linewidth}
    \centering
    \includegraphics[height=14.7em]{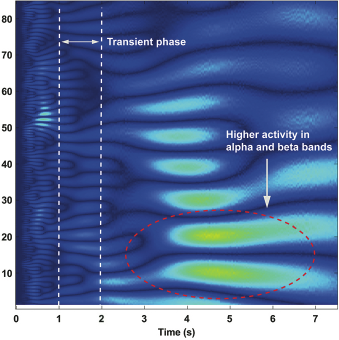}
    \caption{Correct estimation.}
    \label{fig:accurate_estimation}
  \end{subfigure}
  \begin{subfigure}[t]{0.32\linewidth}
    \centering
    \includegraphics[height=14.7em]{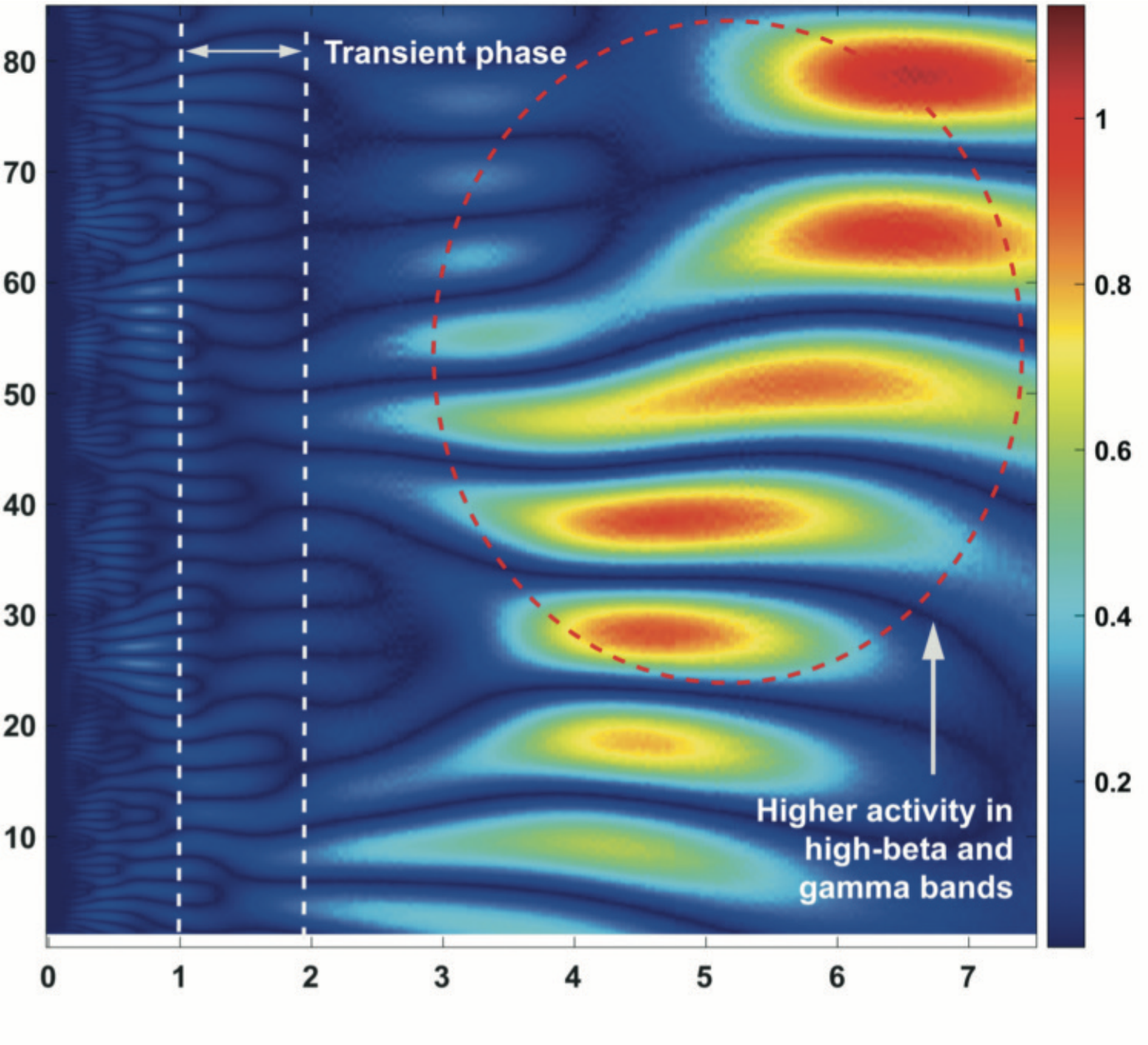}
    \caption{Underestimation.}
    \label{fig:underestimation}
  \end{subfigure}
  \vspace{-0.95em}
  \caption{Time-frequency analysis of EEG signals recorded during 6-second events for over-, under-, and correct estimation.}
  \label{fig:teaser}  
}

\abstract{
Achieving a high level of immersion and adaptation in virtual reality (VR) requires precise measurement and representation of user state.
While extrinsic physical characteristics such as locomotion and pose can be accurately tracked in real-time, reliably capturing mental states is more challenging.
Quantitative psychology allows considering more intrinsic features like emotion, attention, or cognitive load.
Time perception, in particular, is strongly tied to users' mental states, including stress, focus, and boredom.
However, research on objectively measuring the pace at which we perceive the passage of time is scarce.
In this work, we investigate the potential of electroencephalography (EEG) as an objective measure of time perception in VR, exploring neural correlates with oscillatory responses and time-frequency analysis.
To this end, we implemented a variety of time perception modulators in VR, collected EEG recordings, and labeled them with overestimation, correct estimation, and underestimation time perception states.
We found clear EEG spectral signatures for these three states, that are persistent across individuals, modulators, and modulation duration.
These signatures can be integrated and applied to monitor and actively influence time perception in VR, allowing the virtual environment to be purposefully adapted to the individual to increase immersion further and improve user experience.
A free copy of this paper and all supplemental materials are available at \url{https://vrarlab.uni.lu/pub/brain-signatures}.
}

\keywords{User state, time perception, electroencephalography, virtual reality}

\begin{document}

\manuscriptnote{
    This article has been accepted for publication in IEEE Transactions on Visualization and Computer Graphics. 
    This is the author's version which has not been fully edited and content may change prior to final publication.
    Citation information: DOI 10.1109/TVCG.2025.3549570
}

\firstsection{Introduction}
\label{sec:introduction}

\maketitle

User state is critical to creating immersive and adaptive virtual reality (VR) experiences.
Capturing user state in VR occurs on three levels: physical, physiological, and psychological.
The physical tracking of movement or pose in real-time is fundamental to the sense of presence and ensuring sensory contingency between user actions and virtual feedback~\cite{slater1993representation}.
Physiological state, such as pupillometry, body temperature, electrocardiography (ECG), and electroencephalography (EEG), is typically measured using external sensors and allows the monitoring of vital reactions to the virtual environment.
Psychological state, however, such as attention, cognitive load, or stress levels, cannot be recorded directly.
Instead, it is derived from physical and physiological measurements, using quantitative psychology and neural imaging.
Together, these states signify an essential building block for designing adaptive virtual environments that dynamically accommodate and improve user experience.

The way we perceive the passage of time is another mental state that has recently gained attention from researchers in neuroscience and psychology, as well as computer scientists focusing on human-computer interaction (HCI).
Humans perceive time subjectively, with happy or adventurous time seemingly passing faster than sad or idle time.
Different and often interwoven factors affect the pace at which we perceive the passage of time. 
For example, emotions~\cite{angrilli1997influence, droit2007emotions, smith2011effects}, cognitive engagement~\cite{block2010impact, block2010cognitive, hicks1976prospective}, and novelty~\cite{matthews2016repetition, ulrich2006perceived, utegaliyev2024expected} can expand or compress our temporal judgments.
This plasticity of time perception is a valuable asset for designing adaptive virtual environments and VR experiences, as it strongly correlates with other psychological states, such as stress, boredom, and focus~\cite{botev2021chronopilot, igarzabal2021happens, morin2024mindfulness}.

By modulating time perception in VR, we can also enhance general well-being and productivity in collaborative environments, for example, by synchronizing the paces across team members~\cite{botev2021chronopilot}.
However, leveraging time perception as a psychological user state in VR presents two challenges.
First, achieving precise and purposeful control over how users perceive time requires a comprehensive and systematic study of time perception in VR.
This involves both investigating classical time perception modulators in VR and exploring VR itself as a time perception modulator, as time compression has often been reported during the use of VR technologies~\cite{mullen2021time, read2021engagement, schneider2011effect}.
Second, as discussed before, time perception is a psychological state and thus only accessible in real-time through adequate physiological state measurements.
Current methods for measuring time perception rely on subjective and offline self-report questionnaires, yet a real-time monitoring and modulation loop in VR calls for an objective, online, and physiology-based measurement method.

This paper constitutes a first step towards addressing these challenges.
We implemented a variety of classic time perception modulators in VR, in both visual and auditory modalities, and collected an extensive dataset of users' interval timing.
Moreover, we investigated EEG's potential as an objective and online measure of time perception in VR.
EEG is an excellent candidate for time perception measurement as a non-invasive neuroimaging method with high temporal resolution.
Furthermore, modern EEG devices are affordable and portable, allowing seamless coupling with other wearable technologies, such as VR headsets.
To this end, we recorded participants' EEG signals during exposure to time perception modulators and labeled them according with three time perception states: \emph{overestimation}, where participants estimated a duration longer than its objective temporal length; \emph{correct estimation}, where participants estimated a duration very close to its objective temporal length; and \emph{underestimation}, where participants estimated a duration shorter than its objective temporal length.
We then grouped the collected data by time perception states and performed EEG analyses to detect possible corresponding signatures.
We identified clear spectral signatures for the different time perception states that persist across individuals, modulators, and modulation duration.

\subsection{Related Work}
\label{subsec:related_work}

Study of time perception in VR began with the reports of its effect in compressing perceived durations~\cite{mullen2021time, read2021engagement, schneider2011effect}.
This effect is especially prominent while playing video games, leading to the phenomenon of \emph{time loss}, where players in virtual environments lose track of real-world time~\cite{nuyens2020potential, rau2006time, triberti2018matters, wood2007time}.
Several studies have investigated and invested in this phenomenon as a therapeutic resource for patients undergoing difficult treatments~\cite{chirico2016elapsed, schneider2011effect}.

With its growing availability and affordability, VR technology has captured significant attention from the scientific community, as a unique testbed for researching and modulating perception, including the perception of time.
In most studies that explored time perception in VR, the virtual environment is used merely as a basic platform for conducting the experiment.
However, in recent years, a growing number of studies have begun to explore VR itself as a focus of investigation in relation to time perception.\cite{de2023altering}

In one of the earliest studies, Bruder and Steinicke~\cite{bruder2014time} investigated time perception modulation during walking in a virtual environment compared to a real environment.
Similarly, Lugrin et al.~\cite{lugrin2019experiencing} asked participants to estimate the duration of their waiting time in both a real room and its 3D replica in VR.
Mullen and Davidenko~\cite{mullen2021time} conducted an intriguing study to monitor time perception in VR and non-VR setups under otherwise identical conditions.
Schatzschneider et al.~\cite{schatzschneider2016turned} and Fischer et al.~\cite{fischer2022time} tested the relevance of real-world \emph{zeitgebers} simulated within VR environments.
Malpica et al.~\cite{malpica2022larger} explored the transferability of multiple \emph{magnitude} time perception modulators in VR compared to conventional displays.
However, since the 2020s, there has been a notable rise in studies on time perception in VR, with a shift of focus from psychology to VR-native concepts such as embodiment~\cite{lugrin2019experiencing, unruh2021influence, unruh2023body}, user interaction~\cite{read2021engagement, read2023influence, unruh2023body}, and environmental dynamics~\cite{bogon2024age, read2021engagement, read2023influence}.

As a non-invasive neuroimaging method with high temporal resolution, EEG has consistently been a compelling choice for studying time perception at the neural level.
EEG enables the monitoring of neural rhythmic electrical activity, categorized into standard frequency ranges that have been extensively studied and linked to distinct mental states and cognitive processes.
In recent years, several studies tested EEG for identifying neural signatures of time perception states.
For example, Ernst et al.~\cite{ernst2017p3} explored the potential of EEG to predict the overestimation of sub-second intervals caused by the \emph{oddball} effect.
Their study investigates the significance of Event-Related Potentials (ERPs), which are brief, time-locked changes in neural activity triggered by specific sensory, cognitive, or motor events.
In another work using EEG, Ghaderi et al.~\cite{ghaderi2018time} investigated the differences between the over- and underestimation states during a 15-minute mindfulness task.
Grabot et al.~\cite{grabot2019strength} studied the EEG signatures associated with accurate time perception.
Horr et al.~\cite{horr2016perceived} conducted a study using sub-second interval comparison tasks to examine neural activity during overestimation compared to underestimation states.

In the last few years, several studies have employed EEG to investigate time perception in VR, reflecting a growing interest in integrating objective and real-time measurements of time perception.
Li and Kim~\cite{li2021effect} used EEG to monitor neural activity while exploring the effect of cognitive load on time perception.
Martins e Silva et al.~\cite{martins2022non} monitored changes in cortical activity, as well as users' performance on a timing task under the effect of different levels of motion speed in VR.
Finally, Moinnereau et al.~\cite{moinnereau2023quantifying} recruited eight gamers and recorded their EEG signals, along with ECG and electrooculography (EOG), while they played action games in VR.
They then mapped the data to the gamers' reported time perception and time loss, identifying features predictive of time perception.
Though few, these studies already hint at the potential of EEG to objectively measure time perception in virtual environments and its promise to enhance the immersion and adaptability of virtual experiences.

\section{Methodology}
\label{sec:methodology}
We tested 15 different modes of 5 time perception modulators, 
including \emph{emotion}~\cite{angrilli1997influence, droit2010time, droit2007emotions, smith2011effects}, \emph{cognitive load}~\cite{block2010impact, block2010cognitive, hicks1976prospective}, \emph{oddball}~\cite{schweitzer2017associated, simchy2018expectation, tachmatzidou2023attention, ulrich2006perceived}, \emph{magnitude}~\cite{fabbri2012theory, xuan2007larger}, and \emph{expectation}~\cite{harrison2007rethinking, kim2017effect, kuroki2015manipulating, soderstrom2018users}, along with a \emph{control} condition with no modulator.
All modes were presented three times as events of length 2, 4, and 6 seconds, 
except for the three modes of \emph{oddball} modulator (cf.~\cref{subsec:modulators}), resulting in a total of 42 events.
The choice of these specific durations is inspired by the work of Angrilli et al.~\cite{angrilli1997influence}, aiming to maintain a tolerable duration for the VR experience while providing the opportunity to incorporate a significant number of observations for a comprehensive analysis.
Participants were asked to estimate the duration of each event in seconds. 
All events were presented in a randomized order.

\subsection{Procedure}
\label{subsec:procedure}
\begin{figure}[h!]
  \centering
  \includegraphics[width=\columnwidth]{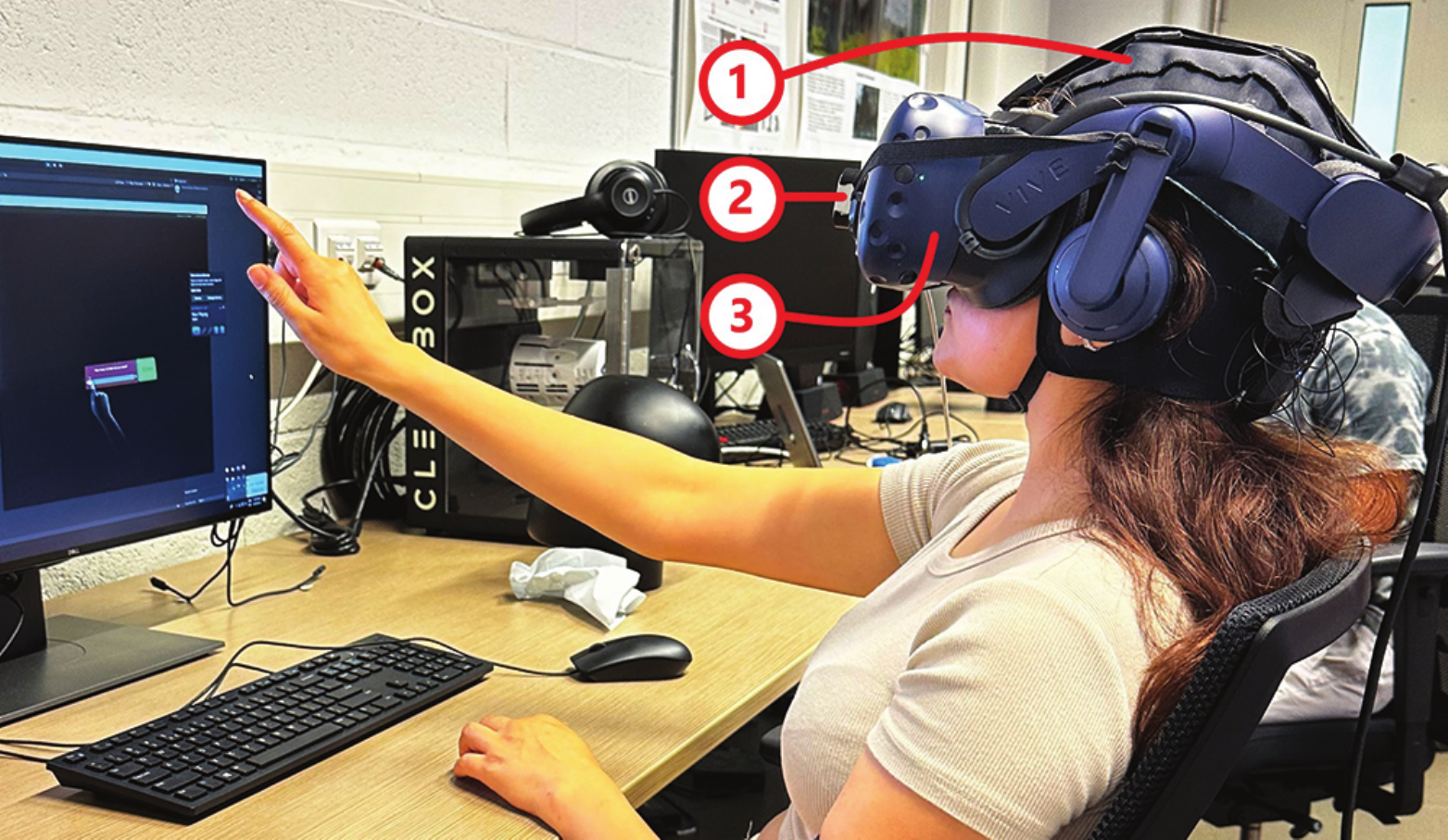}
  \caption{
    Experiment setup showing (1)~EEG cap, (2)~headset-mounted hand-tracking device, and (3)~VR headset.
  }
  \label{fig:setup}
\end{figure}

At the beginning, participants wore a Unicorn Hybrid Black EEG cap with 8 channels\footnote{\url{https://www.unicorn-bi.com/}}.
Although an EEG device with more electrodes could potentially produce more accurate results, the simplicity of the 8-channel headset offered practical advantages, including shorter setup times, reduced experiment duration, and improved participant comfort, making it well-suited for our study.
On top of the EEG cap, participants wore an HTC Vive Pro Eye VR headset\footnote{\url{https://business.vive.com/sea/product/vive-pro-eye/}} as shown in~\cref{fig:setup}.

Participants were instructed to sit in a comfortable position and minimize body and head movement as much as possible during the experiment.
Participants first completed a brief practice session in a VR environment similar to the one used in the actual experiment.
During the practice session, they had the opportunity to work and get familiar with the VR user interface and the timing task by estimating the duration of several practice events.
Participants were instructed to refrain from counting the seconds or use any time-keeping techniques during the practice session or the experiment.
In the practice session, participants received feedback showing the objective duration of the events following their estimations.
This feedback was provided to serve as a time perception calibration.

\begin{figure}[h!]
    \centering
    \begin{subfigure}[t]{0.49\columnwidth}
        \centering
        \includegraphics[width=\textwidth]{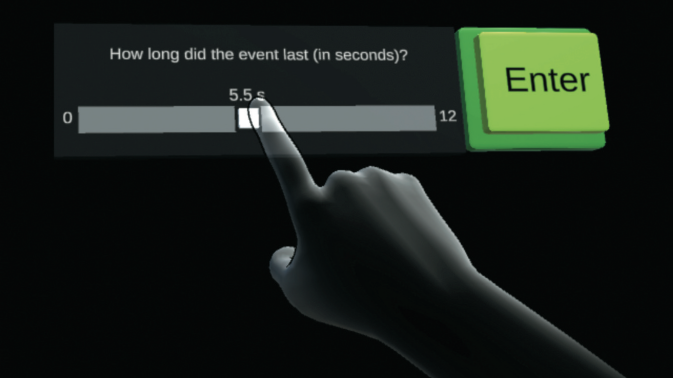}
        \caption{Timing task slider asking participants: \emph{How long did the event last (in seconds)?}}
        \label{fig:ui_timing}
    \end{subfigure}
    \hfill
    \begin{subfigure}[t]{0.49\columnwidth}
        \centering
        \includegraphics[width=\textwidth]{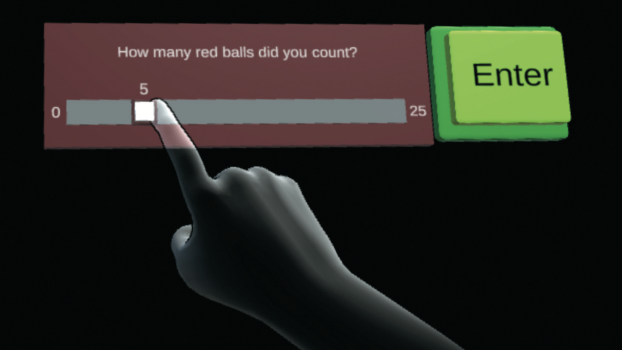}
        \caption{Cognitive task slider asking participants: \emph{How many red balls did you count?}}
        \label{fig:ui_cognitive}
    \end{subfigure}    
    \caption{Gesture-based user interface used in our study.}
    \label{fig:ui}
\end{figure}

During the experiment, participants estimated the duration of 42 events presented in random order.
Each event was preceded by a 3-second resting pause on a black screen, which was also used for baseline correction for the EEG recordings.
Each event started with a monotonous notification tone as the prompt and ended with the appearance of a time estimation slider (\cref{fig:ui_timing}).

The estimation slider covers a range of 0 to 12 seconds, twice as long as the study's longest event, to reduce the central tendency bias~\cite{xiang2021confidence}.
Revisiting the work of Angrilli et al.~\cite{angrilli1997influence}, we set the resolution (steps) of the estimation slider to 0.1\,s, resulting in 120 evenly spaced steps along the slider.
Due to the large number of steps and the constraints that ruled out a long slider--which would require head rotation and potentially cause EEG noise--we opted for mid-air hand-tracking to enable natural and smooth adjustments on a shorter slider.
Hand-tracking was implemented using a Leap Motion Controller\footnote{\url{https://www.ultraleap.com/product/}} mounted on the VR headset.

\subsection{Modulators}
\label{subsec:modulators}

\begin{figure}[h!]
  \centering
  \includegraphics[width=\columnwidth]{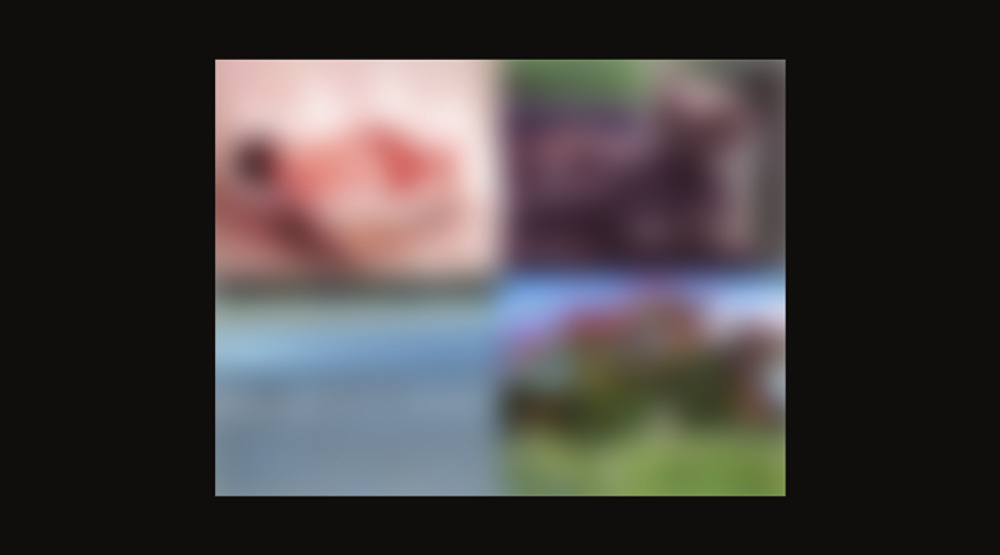}
  \caption{
    Example collage of four pictures for each mode of emotion modulator; the pictures are blurred due to the authors of the image datasets not allowing publication.
  }
  \label{fig:emotions}
\end{figure}

\paragraph{Emotion}
For the \emph{emotion} modulator, we implemented four visual modes:
positive valence and high arousal, positive valence and low arousal, negative valence and low arousal, and neutral.\footnote{
    These modes are actually appraisals of \emph{affect},
    but since many authors refer to them as emotions, 
    we decided to name this modulator ``emotion modulator'' instead of ``affect modulator''.
}
The negative valence and high arousal mode was not included based on the negative feedback we received during the pilot study to avoid potential emotional distress.
For each mode, we exposed the participants to a collage of four pictures against a black screen to induce emotional responses (\cref{fig:emotions}).
The decision to present four pictures instead of one was made to minimize the chances of participants having a personal emotional reaction to a specific triggering object in one picture.
The images were sourced from the Geneva Affective Picture Database (GAPED)~\cite{dan2011geneva} and the Nencki Affective Picture System (NAPS) database~\cite{marchewka2014nencki, riegel2016characterization}, which include arousal and valence ratings.
Within this modulator, the mode labeled as 'neutral' is the intra-modulator control condition.
We selected these datasets from a pool of five standard and top-cited affective image sources, based on dataset size and accessibility.
The images were chosen based on the datasets' rating systems, with detailed values accessible in this study's \hyperref[sec:supplementary_materials]{supplementary materials}.

\begin{figure*}[ht!]
    \centering
    \begin{subfigure}[t]{0.32\textwidth}
        \centering
        \includegraphics[width=\textwidth]{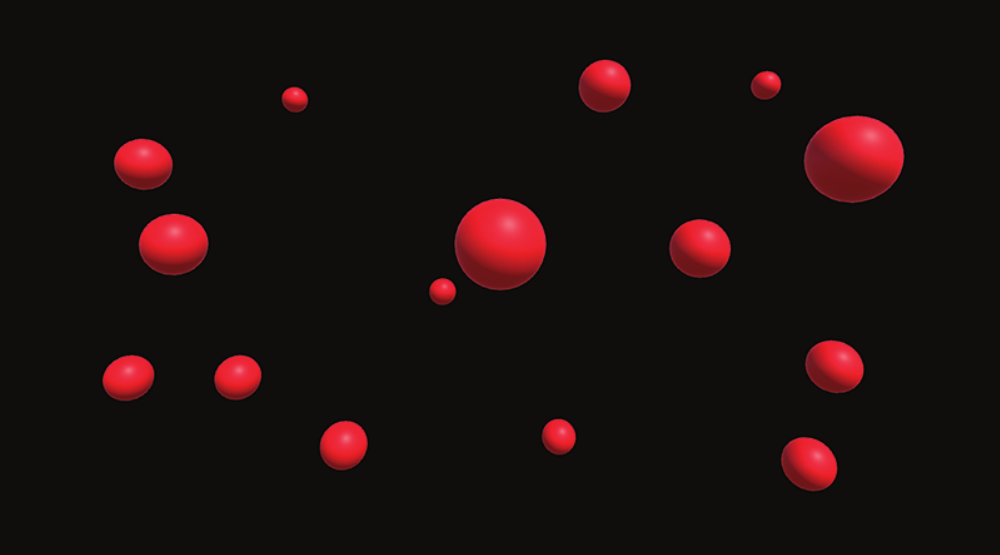}
        \caption{Low cognitive load.}
        \label{fig:cognitive-load_low}
    \end{subfigure}
    \hfill
    \begin{subfigure}[t]{0.32\textwidth}
        \centering
        \includegraphics[width=\textwidth]{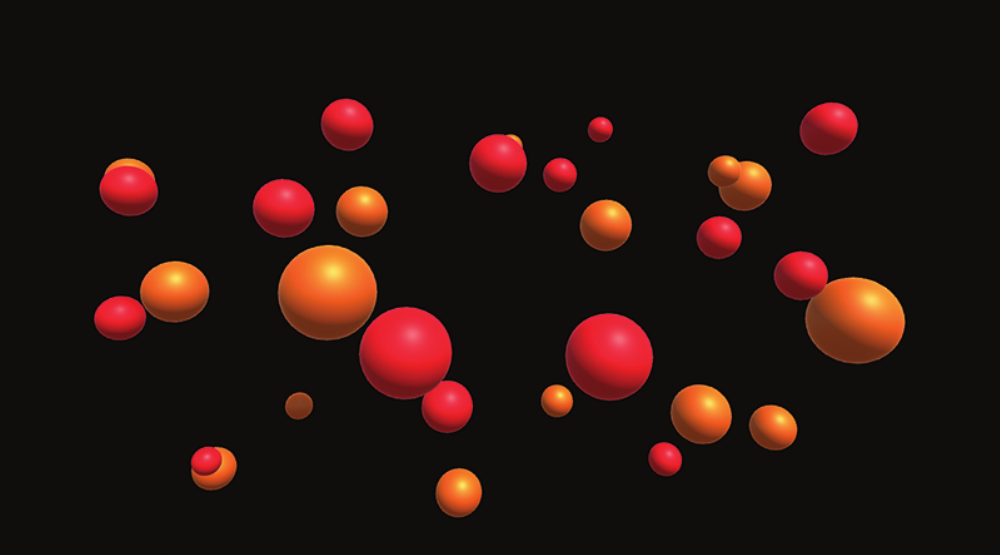}
        \caption{Medium cognitive load.}
        \label{fig:cognitive-load_medium}
    \end{subfigure}
    \hfill
    \begin{subfigure}[t]{0.32\textwidth}
        \centering
        \includegraphics[width=\textwidth]{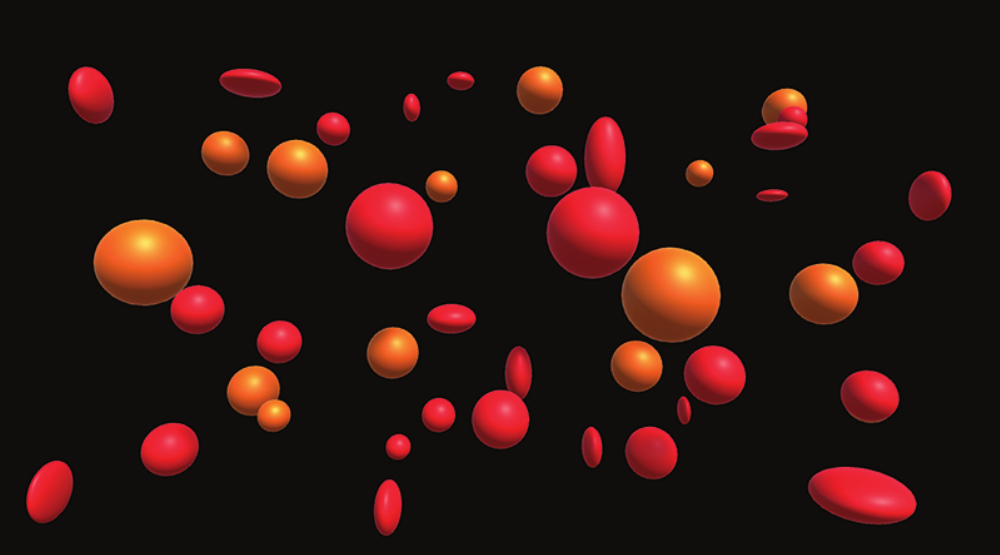}
        \caption{High cognitive load.}
        \label{fig:cognitive-load_high}
    \end{subfigure}    
    \caption{Three modes of visual cognitive modulator.}
    \label{fig:cognitive_load}
\end{figure*}

\paragraph{Cognitive Load}
We implemented three visual modes: low, medium, and high cognitive load.
For this modulator, participants were presented with a set of three-dimensional objects against a black screen and were instructed to count the number of red balls within the set (\cref{fig:cognitive_load}) as the cognitive task.
In the low cognitive load mode, only red balls are displayed, simplifying the counting task.
However, in the medium and high modes, we added distractions in the form of orange balls and red spheroids, requiring participants to discriminate colors and shapes before counting.
This layered design for the cognitive task was inspired by Hicks et al.~\cite{hicks1976prospective}, who used a card-sorting task to study the effect of information processing on time perception.
For this modulator, the mode medium is considered the intra-modulator control mode.
A similar slider to the estimation slider was designed for the participants to enter their answers to the cognitive task (\cref{fig:ui_cognitive}).
This slider appeared right after the participants entered their duration estimation.
The cognitive task's slider was set to a range of 0--25 and had a different color to avoid confusion.

\begin{figure}[h!]
    \centering
    \begin{subfigure}[t]{0.3\columnwidth}
        \centering
        \includegraphics[width=8em]{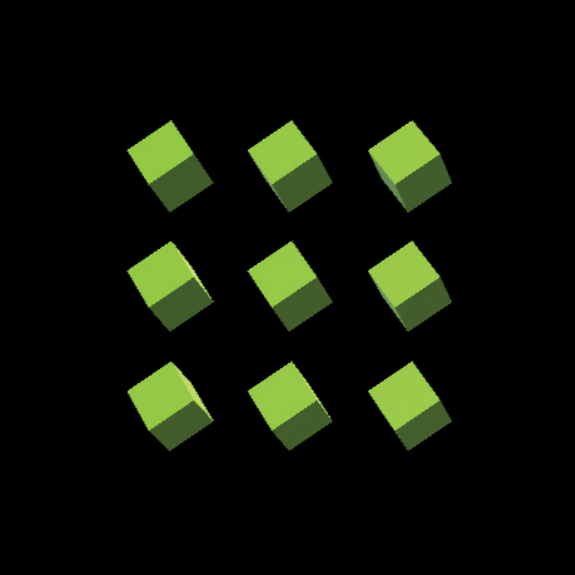}
        \caption{No-oddball.}
        \label{fig:oddball_no-oddball}
    \end{subfigure}
    \hfill
    \begin{subfigure}[t]{0.3\columnwidth}
        \centering
        \includegraphics[width=8em]{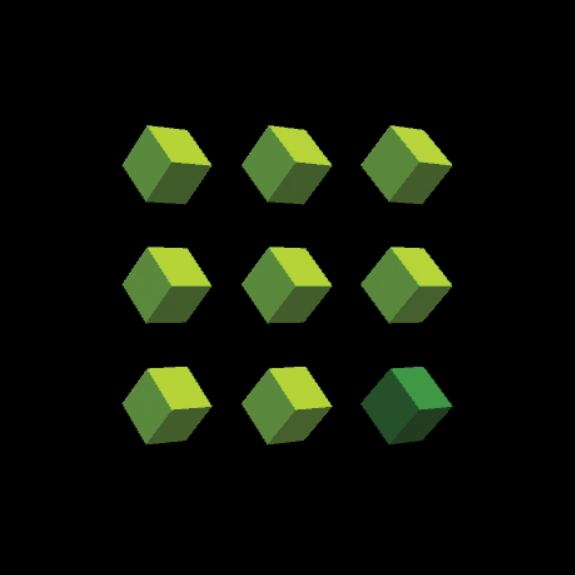}
        \caption{Context-relevant.}
        \label{fig:oddball_context-relevant}
    \end{subfigure}
    \hfill
    \begin{subfigure}[t]{0.3\columnwidth}
        \centering
        \includegraphics[width=8em]{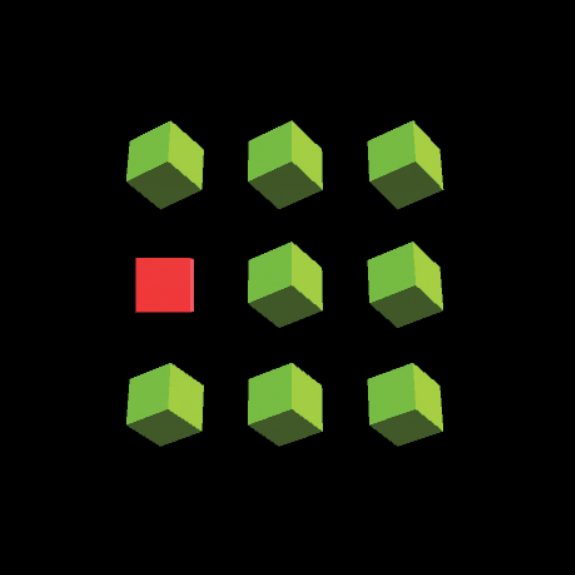}
        \caption{Context-irrelevant.}
        \label{fig:oddball_context-irrelevant}
    \end{subfigure}    
    \caption{Three modes of visual oddball modulator.}
    \label{fig:oddball}
\end{figure}

\paragraph{Oddball}
We implemented three visual modes: no-oddball, context-relevant, and context-irrelevant oddball~\cite{schweitzer2017associated}.
We designed this modulator using a configuration of nine rotating green cubes arranged in three rows against a black screen (\cref{fig:oddball}).
In the no-oddball mode, all cubes rotate synchronously and share the same color, whereas in the context-relevant mode, one of the cubes has a darker shade of green, and in the context-irrelevant mode, one is red and static.
For this modulator, however, the modes were presented only once, for 4 seconds, to preserve the surprise effect.
The no-oddball mode serves as the intra-modulator control condition for this modulator.

\paragraph{Magnitude}
The magnitude time perception modulator is better known in visual modes, where an object's appearance seems to last longer with an increased size.
However, many studies have shown that the effect of magnitude on time perception extends to other modalities and causes overestimations for heavier, larger, louder, more, faster, and brighter stimuli~\cite{fabbri2012theory, kaneko2009perceived, lu2011weight, xuan2007larger}.
Taking advantage of this property, we designed the magnitude modulator in the auditory modality.
We cut audio clips from the beginning of Barack Obama's speech at the White House Correspondents' Dinner in 2009, and by changing the standard sampling rate from 44,100\,Hz to 88,200\,Hz and 22,050\,Hz, we created three modes for this modulator: natural, high, and low pitch.
Participants were presented with a black screen and no visual stimulus while listening to the audio clips.
The natural mode was designed to act as the intra-modulator control mode in this modulator.

\paragraph{Expectation}
For this modulator, we implemented two visual modes: slow and fast.
An animated progress bar was presented against a black screen, starting at 10\% complete.
In the slow mode, the bar starts advancing 0.36\% per second, while in the fast mode, the progress rate is 5\% per second.
We did not implement an intra-modulator control condition for this modulator, considering the short duration of the events could make the difference between the treatment and control modes imperceptible.

\paragraph{Control}
The control events consisted of a black screen with no visual or auditory stimuli and were implemented as inter-modulator control condition.

\subsection{Sample Size and Recruitment}
\label{subsec:sampleSize}
We conducted an \emph{a priori} calculation for sample size using G*Power~\cite{faul2007g} with a repeated measures analysis of variance (RM-ANOVA), setting the effect size to 0.25, significance level to ($\alpha = 0.05$), and aiming for a statistical power of $1-\beta = 0.95$), with the EEG signal as the dependent variable, repeatedly measured under three conditions: overestimation, correct estimation, and underestimation, as the independent variables.
We then fitted a normal distribution of different time perception states and, based on the work of Angrilli et al.~\cite{angrilli1997influence}, estimated 8 correct estimations out of 42 events per participant, 
accounting for the smallest number of measurements per condition.
This resulted in a minimum of 30 participants as the target sample size.

Next, participants were recruited by posting flyers on campus and in nearby social venues, as well as distributing information through university email lists.
The inclusion criteria were normal or corrected-to-normal vision and proficiency in English, while the exclusion criteria included a history of neurological disorders and susceptibility to cybersickness.
A total of 36 participants (12 identifying as female, 24 identifying as male) aged 19 to 43 years (Mean=28.7, SD=5.7) were recruited.
We had to exclude data from three individuals: two did not follow the instructions correctly, and a third one disclosed an acute neurological condition after the experiment. 
Participants were informed both verbally and in written form of their right to ask questions, interrupt or leave the study at any time, and withdraw their consent without facing any consequences.
Written informed consent was obtained from all participants prior to their participation in the experiment, and they were provided with a debriefing session explaining the study's objectives after the experiment. 
The entire procedure, including setting up the VR and EEG headsets, took approximately 45 minutes to complete.
The actual experiment in VR lasted about 20 minutes.
Upon completion, the participants were compensated with a gift card worth 15 euros. 
The study received ethical approval from the Ethics Review Panel of the University of Luxembourg, under approval number ERP 22-031 BANANA.

\begin{figure}[h!]
  \centering
  \includegraphics[width=.5\columnwidth]{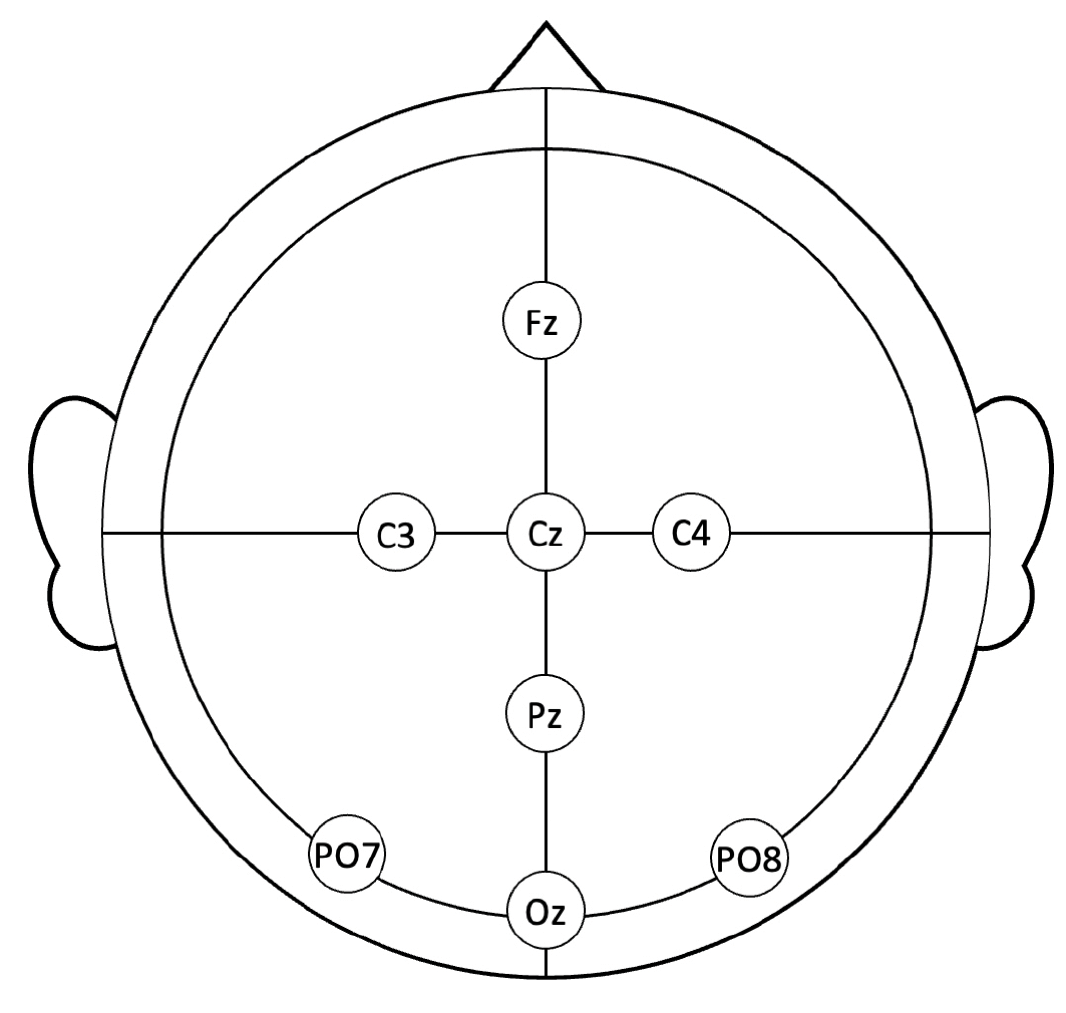}
  \caption{
    EEG channel locations of the EEG device.
  }
  \label{fig:eeg_channels}
\end{figure}

\subsection{Data Collection and Processing}
\label{subsec:data}
EEG data were continuously recorded during the experiment through eight channels (Fz, C3, Cz, C4, Pz, PO7, Oz, and PO8) 
with a sampling rate of 250\,Hz according to the international 10--20 EEG electrode system (\cref{fig:eeg_channels}).
The reference and ground electrodes were placed on the right and left mastoid bones.
Conductive gel was applied during electrode placement to ensure impedance matching.
Power line interference was removed from the EEG data using a 50\,Hz notch filter.

For preprocessing EEG signals, we used the MATLAB-EEGLAB toolbox~\cite{delorme2004eeglab}.
The EEG data were re-referenced using average mastoids to minimize the effects of common noise sources.
We employed a fourth-order Butterworth Finite Impulse Response (FIR) band-pass filter to isolate the frequency components of interest.
The high-pass cutoff frequency was set at 0.05\,Hz to eliminate slow drifts and baseline shifts, while the low-pass cutoff frequency was set at 80\,Hz to exclude high-frequency noise.
Furthermore, Independent Component Analysis (ICA) was applied to remove artifacts caused by eye movements, blinking, muscle activity, and other non-neural sources.

\begin{figure*}
  \centering
  \includegraphics[width=0.7\linewidth]{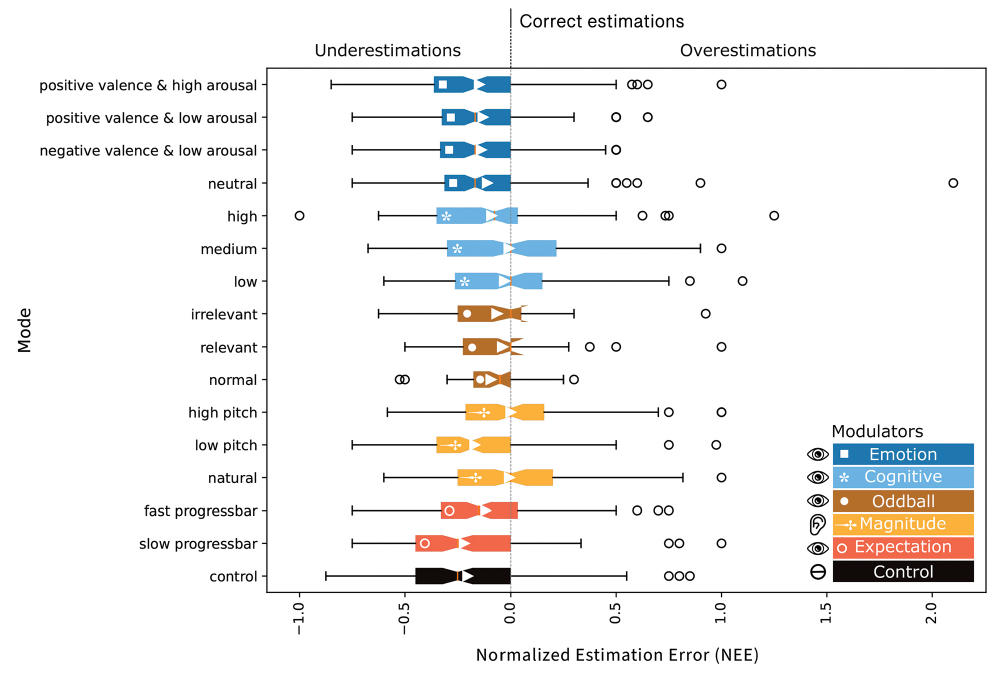}
  \caption{
    Normalized perception deviation values for the experiment modulation modes.
  }
  \label{fig:modes_boxplots}
\end{figure*}

We then marked the EEG signals using the timestamps indicating the beginning and end of the experiment's events.
Finally, EEG samples were extracted, spanning a duration of 1\,second before and after each event.
The 1\,second preceding the event was used for the baseline correction, and the 1\,second after the event was included to cover possible lasting modulation of neural activities after the events.
Based on the participants' estimation, the corresponding EEG sample was labeled as either \emph{overestimation}, \emph{correct estimation}, or \emph{underestimation}.
The term \emph{correct estimation} reflects a margin of error of 0.1\,seconds to account for potential inaccuracies in working with the closely spaced steps of the user interface and unfamiliarity with hand-tracking.
This specific margin was set experimentally, based on the feedback we received during the pilot test.
Eventually, we gathered an EEG dataset with 1386 samples from 33 participants, including 337 \emph{overestimation}, 263 \emph{correct estimation}, and 786 \emph{underestimation} samples (\cref{fig:modes_boxplots}).

While analyzing EEG samples, we focused on finding the event-related oscillatory responses for each time perception state in \emph{delta} (0.5–4 Hz), \emph{theta} (4–8 Hz), \emph{alpha} (8–12 Hz), \emph{beta} (13–30 Hz), \emph{high-beta} (20-30 Hz), and \emph{gamma} ($>$30 Hz) frequency bands.
To find potential EEG signatures corresponding to time perception states, we grouped the datasets by the labels \emph{overestimation}, \emph{correct estimation}, and \emph{underestimation}.
We then calculated the oscillatory responses for each of these three states using the Welch-based Power Spectral Density (PSD) method, which transforms a time-domain signal into the frequency domain and estimates its power distribution across the frequency spectrum.
A topographical analysis was further performed to capture the spatial distribution of neural activity across the scalp, highlighting possible localized patterns.
In addition, we examined the consistency of the EEG signatures across individuals by grouping the samples by participants and calculating the signal-to-noise ratio (SNR) for each frequency band per participant-state.
Then, we investigated the signatures grouped by modulation duration and investigated PSD per duration-state.
This was followed by a time-frequency analysis using the Morlet Wavelet Transform (MWT), characterizing the dynamic evolution of neural oscillations across time and frequency to more closely examine the special case of 2-second events.

\section{Results}
\label{sec:result}

\subsection{Time Perception Modulators}
\label{subsec:TPModulators}
To normalize samples for the varying duration of the events, we calculated the \emph{Normalized Estimation Error} (NEE) using \cref{eq:NEE}, such that a positive value means overestimation, $0$, correct estimation, and underestimation returns negative NEE.

\begin{equation}
\text{NEE} = \frac{\text{Estimation} - \text{Objective\ duration}}{\text{Objective\ duration}}
\label{eq:NEE}
\end{equation}

We then investigated the potential influence of fatigue and possible learning effects on participants' estimations by calculating Pearson’s and Spearman’s correlation coefficients between the elapsed time from the start of the experiment and the normalized perception deviation.
The Pearson correlation coefficient was found to be 0.025 ($p = .3521$), and the Spearman correlation coefficient was 0.016 ($p = .5486$), denoting thus no correlation. Furthermore, we calculated the correlation between the duration of the events and NEE.
A correlation coefficient of -0.15 ($p < .0001$) was obtained with the Pearson and -0.12 ($p < .0001$) with the Spearman method, indicating a weak negative linear relationship between the duration of the stimuli and NEE.

We also performed $t$-tests to compare the normalized deviation of each mode with its intra- and inter-modulator control modes (\cref{tab:ttest_modes}).
The first three modes of emotion modulator, including positive valence and both high and low arousal and negative valence and low arousal, did not show a significant impact of time perception modulation in contrast to the neutral mode, which was implemented as the intra-modulator control condition.
Nor did these conditions confirm a significant modulation compared to the experiment inter-modulator control modes.
However, when compared with the inter-modulator control, the neutral mode proved significantly effective in expanding the perception of the durations (\cref{tab:ttest_modes}: Emotion modulator).

Similarly, compared to the inter-modulator control events, all modes of cognitive and oddball modulators significantly slowed subjective time for the participants, with cognitive load demonstrating a more pronounced effect.
Nevertheless, no significant difference was observed between these modes and their corresponding intra-modulator control modes (\cref{tab:ttest_modes}: Cognitive and Oddball modulators).

As for the magnitude modulator, the low pitch significantly compressed perceived durations, in contrast with the intra-modulator mode, the natural pitch.
Both high and natural pitch modes expanded subjective time compared to the inter-modulator control modes (\cref{tab:ttest_modes}: Magnitude modulators).

\begin{figure*}[h!]
  \centering
  \includegraphics[width=\linewidth]{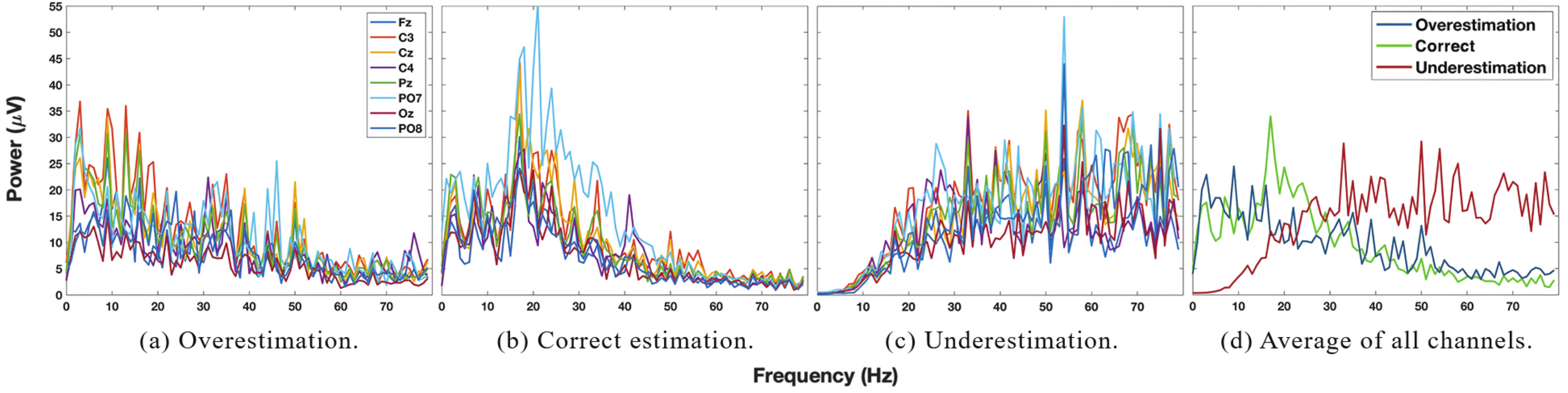}
  \caption{
    Spectral activity of EEG channels averaged across all participants for (a)~overestimation, (b)~correct estimation, and (c)~underestimation states, and (d)~spectral activity averaged across all channels and participants.
  }
  \label{fig:PSD}
\end{figure*}

In the exception modulator case, the fast and slow progress bar modes demonstrated significantly different impacts on time perception.
Compared to the slow progress bar and inter-modulator control condition, the fast progress bar significantly expanded the perception of duration (\cref{tab:ttest_modes}: Expectation modulators).

\begin{table}[h!]
\caption{Pairwise comparison of normalized estimation errors for every mode against intra- and inter-modulator control events.}
\vspace{-1mm}
\centering
\small
\setlength{\tabcolsep}{4pt}
\begin{tabular*}{\columnwidth}{@{\extracolsep{\fill}}p{0.5cm}cp{0.05cm}p{1.2cm}ccc}
\toprule
\textbf{Modulator} & \textbf{Mode} & \multicolumn{2}{l}{\textbf{Control}} & $\textbf{t-value}$ & $\textbf{p-value}$ & $\textbf{Cohen's d}$\\
\midrule

\multirow{7}{*}{\rotatebox[origin=c]{90}{\textbf{Emotion}}}
    & +High & \multirow{3}{*}{\rotatebox[origin=c]{90}{\textbf{intra}}} & Neutral & $-1.00$ & $.3179$ & $-0.11$\\
    & +Low & & Neutral & $-0.56$ & $.5734$ & $-0.06$ \\
    & -Low & & Neutral & $-0.77$ & $.4411$ & $-0.09$ \\
\cmidrule(lr){2-7}
    & +High & \multirow{4}{*}{\rotatebox[origin=c]{90}{\textbf{inter}}} & Control & $\ \ \ 1.44$ & $.1532$ & $\ \ \ 0.17$ \\
    & +Low & & Control & $\ \ \ 1.70$ & $.0919$ & $\ \ \ 0.24$ \\
    & -Low & & Control & $\ \ \ 1.56$ & $.1211$ & $\ \ \ 0.22$ \\
    & \ Neutral & & Control & $\ \ \ 2.19$ & $\textbf{.0310}$ & $\ \ \ 0.27$\\
\midrule
\multirow{5}{*}{\rotatebox[origin=c]{90}{\textbf{Cognitive}}}
    & High & \multirow{2}{*}{\rotatebox[origin=c]{90}{\textbf{intra}}} & Medium & $-1.83$ & $.0626$ & $-0.22$\\
    & Low & & Medium & $-0.46$ & $.6492$ & $-0.06$ \\
\cmidrule(lr){2-7}
    & High & \multirow{3}{*}{\rotatebox[origin=c]{90}{\textbf{inter}}} & Control & $\ \ \ 2.68$ & $\mathbf{.0087}$ & $\ \ \ 0.33$ \\
    & Medium & & Control & $\ \ \ 4.51$ & $\mathbf{<.0001}$ & $\ \ \ 0.55$ \\
    & Low & & Control & $\ \ \ 3.96$  & $\mathbf{.0001}$ & $\ \ \ 0.53$ \\
\midrule
\multirow{5}{*}{\rotatebox[origin=c]{90}{\textbf{Oddball}}}
    & Relevant & \multirow{2}{*}{\rotatebox[origin=c]{90}{\textbf{intra}}} & No-oddball & $\ \ \ 1.00$ & $.3246$ & $\ \ \ 0.23$ \\
    & Irrelevant & & No-oddball & $\ \ \ 0.53$ & $.6014$ & $\ \ \ 0.12$ \\
\cmidrule(lr){2-7}
    & Relevant & \multirow{3}{*}{\rotatebox[origin=c]{90}{\textbf{intra}}} & Control & $\ \ \ 2.73$ & $\textbf{.0082}$ & $\ \ \ 0.51$ \\
    & Irrelevant & & Control & $\ \ \ 2.34$ & $\textbf{.0222}$ & $\ \ \ 0.43$ \\
    & No-oddball & & Control & $\ \ \ 2.45$ & $\textbf{.0158}$ & $\ \ \ 0.37$ \\
\midrule
\multirow{5}{*}{\rotatebox[origin=c]{90}{\textbf{Magnitude}}}
    & High & \multirow{2}{*}{\rotatebox[origin=c]{90}{\textbf{intra}}} & Natural & $\ \ \ 0.18$ & $.8599$ & $\ \ \ 0.20$ \\
    & Low & & Natural & $-4.44$ & $\mathbf{<.0001}$ & $-0.53$ \\
\cmidrule(lr){2-7}
    & High & \multirow{3}{*}{\rotatebox[origin=c]{90}{\textbf{inter}}} & Control & $\ \ \ 4.78$ & $\mathbf{<.0001}$ & $\ \ \ 0.63$ \\
    & Low & & Control & $\ \ \ 1.02$ & $.3095$ & $\ \ \ 0.11$ \\
    & Natural & & Control & $\ \ \ 4.12$ & $\mathbf{<.0001}$ & $\ \ \ 0.59$ \\
\midrule
\multirow{3}{*}{\rotatebox[origin=c]{90}{\textbf{Expect.}}}
    & Fast PB & \multicolumn{2}{l}{Slow PB} & $\ \ \ 2.96$ & $\textbf{.0039}$ & $\ \ \ 0.34$ \\
\cmidrule(lr){2-7}
    & Fast PB & \multirow{2}{*}{\rotatebox[origin=c]{90}{\textbf{inter}}} & Control & $\ \ \ 2.03$ & $\textbf{.0455}$ & $\ \ \ 0.28$ \\
    & Slow PB & & Control & $-0.38$ & $.7064$ & $-0.05$ \\
\bottomrule
\end{tabular*}
\label{tab:ttest_modes}
\end{table}

Finally, the result of a one-sample $t$-test on the normalized perception deviation under the inter-modulator control condition against $0$ showed a highly significant time compression effect of the control events without any stimuli ($p < .0001$).

\subsection{Oscillatory Responses}
\label{subsec:oscillatoryresponses}
The PSD analysis highlighted significant variations in band power in different frequency ranges; see \cref{fig:PSD}.
The power spectrum of EEG samples of different time perception states was calculated by averaging the spectral data of each channel for all participants and all events.
This average captures the consistent oscillatory patterns shared across all channels.
Samples of overestimation are characterized by heightened lower band powers, particularly evident in the delta and theta bands (\cref{fig:PSD}a).
\Cref{fig:PSD}b indicates increased alpha, low-beta, and medium-beta activities for correct estimation compared to other band powers.
In the underestimation case, we observe increased higher band powers, specifically within the high-beta and gamma bands (\cref{fig:PSD}c).

An ensemble power spectrum, plotted in \cref{fig:PSD}d, was calculated by averaging data from all EEG channels.
This ensemble serves as a consolidated representation of neural activity.
It presents distinctive frequency power for correct estimation, overestimation, and underestimation states, as well as synchronized oscillatory patterns across multiple brain regions.

\subsection{Topographical Analysis}
\label{subsection:Topoanalysis}
\begin{figure*}
  \centering
  \includegraphics[width=\linewidth]{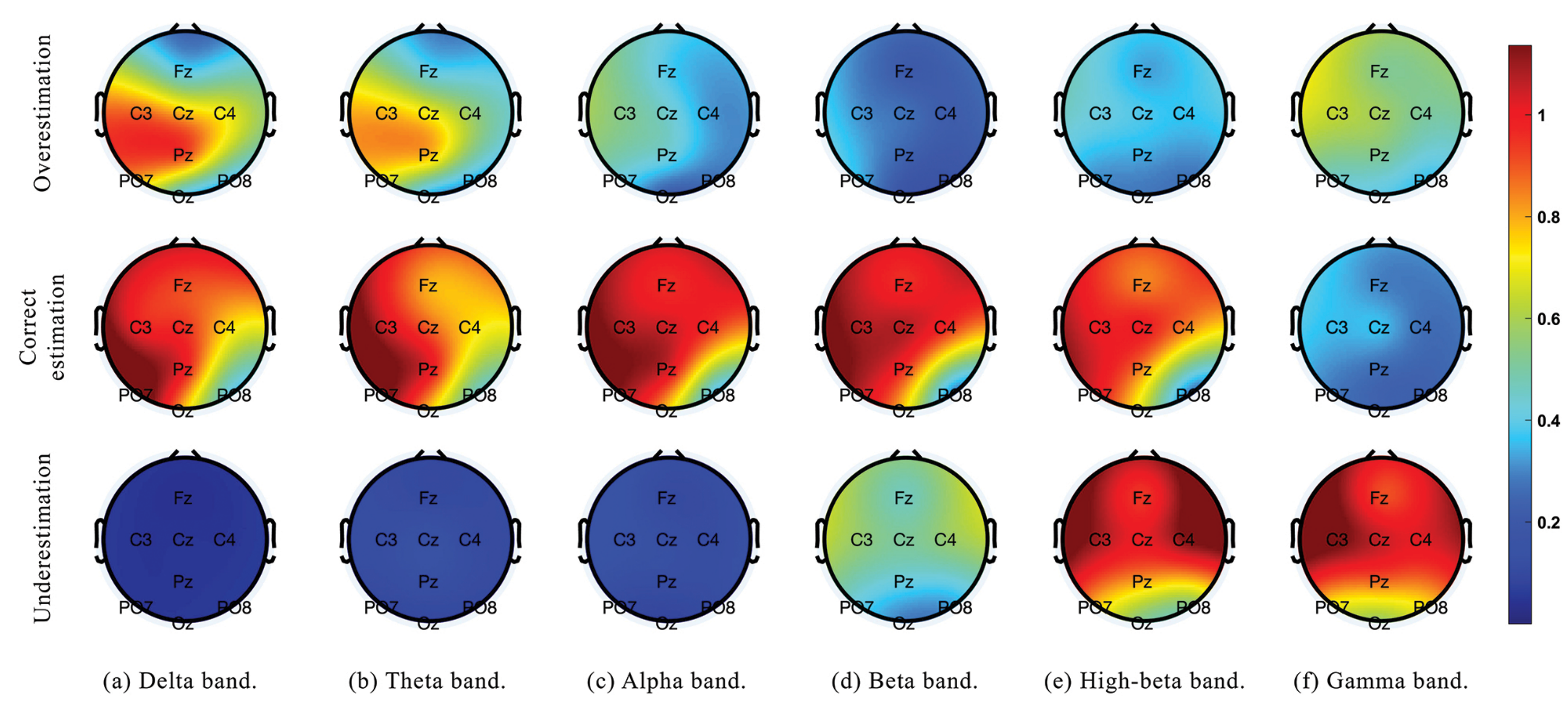}
  \caption{
    Topographic map of different band powers for time perception states.
  }
  \label{fig:topo}
\end{figure*}

We conducted an EEG topographical analysis to explore the spatial distribution of activity in different frequency bands associated with the time perception states.
\Cref{fig:topo} presents the topographical distribution of EEG band powers across the scalp during these three states.

\paragraph{Delta and Theta Bands}
\Cref{fig:topo}, columns a and b, display a consistent topographical pattern characterized by heightened delta and theta band powers during overestimations.
This increased activity is localized in specific electrode channel locations, namely C3, Cz, C4, PO7, and Pz, suggesting a regional specificity in neural processes associated with overestimation.
In contrast, underestimations do not exhibit a strong higher power pattern in the delta and theta bands, indicating a lack of pronounced lower band activity during this state.
The correct estimation case exhibits higher activities in the delta and theta bands and has a strong and consistent signature in Fz, C3, Cz, C4, PO7, Pz, and Oz channels.

\paragraph{Alpha and Beta Bands}
In the analysis of correct estimation, heightened activity was observed in both alpha and beta band power compared to the over- and underestimation states (\cref{fig:topo}c and \cref{fig:topo}d).
In general, correct estimation demonstrates robust activity in all channel locations except PO8, and this activity pattern remains consistent in both the alpha and beta bands.
In contrast, underestimation displays minimal activity in the alpha band, with a subsequent rise in power observed in the beta band (\cref{fig:topo}d), and overestimation activities manifest higher levels in the delta and theta bands (\cref{fig:topo}a and \cref{fig:topo}b) with a diminishing power as the frequency increases in the alpha and beta bands.

\paragraph{High-Beta and Gamma Bands}
\Cref{fig:topo}, columns e and f, present a distinctive topographical pattern for the underestimation state in the high-beta and gamma bands.
Elevated activity in Fz, C3, Cz, C4, and Pz EEG channels in underestimation is evident.
Notably, this pattern contrasts with the lower activity observed in the overestimation state in the high-beta and gamma bands.
The analysis of these two bands for correct estimation state shows a decayed activity from beta to gamma.
Specifically, the diminished power in the gamma band for correct estimation contrasts with the underestimation state.

\paragraph{Time Perception States}
Overall, the activity pattern of correct estimation spans from delta to gamma bands, with medium activity in delta and theta bands, high in alpha and beta bands, followed by a decay in high-beta band and ultimately reaching a low in the gamma band.
The activity pattern of overestimation is high in delta and theta with a subsequent decay in the rest of the bands.
For underestimation, the activity pattern is minimal until the alpha band, and it increases from beta and reaches a peak in the gamma band.

\begin{table}[h!]
\centering
\small
\setlength{\tabcolsep}{4pt}
\caption{ANOVA and pairwise comparisons for EEG bands}
\label{tab:anova_results}
\begin{tabular}{p{1cm} p{2.4cm} p{1.2cm} p{1.2cm} p{1.2cm}}
\toprule
\textbf{Band} & \textbf{Comparison} & \textbf{t-value} & \textbf{p-value} & \textbf{Cohen's d} \\
\midrule
\multirow{3}{*}{\textbf{Delta}} & corr. -- over & $-0.48$ & $\ \ .6369$ & $-0.08$ \\
                                & corr. -- under & $\ \ \ 3.93$  & $\ \ \textbf{.0004}$ & $\ \ \ 0.97$  \\
                                & over \ -- under & $\ \ \ 3.47$  & $\ \ \textbf{.0015}$ & $\ \ \ 0.86$  \\
\cmidrule{2-5}
& \multicolumn{4}{l}{$F(2, 64) = 9.15$, $p = \textbf{.0003}$, $\eta^2 = 0.22$} \\
\midrule
\multirow{3}{*}{\textbf{Theta}} & corr. -- over & $-0.87$ & $\ \ .3886$ & $-0.19$ \\
                                & corr. -- under & $\ \ \ 3.95$  & $\ \ \textbf{.0004}$ & $\ \ \ 1.08$  \\
                                & over \ -- under & $\ \ \ 5.05$  & $<\textbf{.0001}$ & $\ \ \ 1.28$  \\
\cmidrule{2-5}
& \multicolumn{4}{l}{$F(2, 64) = 13.00$, $p < \textbf{.0001}$, $\eta^2 = 0.29$} \\
\midrule
\textbf{Alpha} & \multicolumn{4}{l}{$F(2, 64) = 2.41$, $p = .0977$, $\eta^2 = 0.07$} \\
\midrule
\multirow{3}{*}{\textbf{Beta}}  & corr. -- over & $\ \ \ 3.62$  & $\ \ \textbf{.0010}$ & $\ \ \ 0.74$  \\
                                & corr. -- under & $\ \ \ 0.77$  & $\ \ .4469$ & $\ \ \ 0.17$  \\
                                & over \ -- under & $-2.95$ & $\ \ \textbf{.0058}$ & $-0.73$ \\
\cmidrule{2-5}
& \multicolumn{4}{l}{$F(2, 64) = 6.79$, $p = \textbf{.0021}$, $\eta^2 = 0.18$} \\
\midrule
\multirow{3}{*}{\textbf{High-Beta}} & corr. -- over & $\ \ \ 1.76$  & $\ \ .0879$ & $\ \ \ 0.39$  \\
                                    & corr. -- under & $-1.75$ & $\ \ .0897$ & $-0.42$ \\
                                    & over \ -- under & $-3.55$ & $\ \ \textbf{.0012}$ & $-0.74$ \\
\cmidrule{2-5}
& \multicolumn{4}{l}{$F(2, 64) = 6.16$, $p = \textbf{.0036}$, $\eta^2 = 0.16$} \\
\midrule
\multirow{3}{*}{\textbf{Gamma}} & corr. -- over & $-0.62$ & $\ \ .5371$ & $-0.16$ \\
                                & corr. -- under & $-3.83$ & $\ \ \textbf{.0006}$ & $-0.99$ \\
                                & over \ -- under & $-3.13$ & $\ \ \textbf{.0038}$ & $-0.93$ \\
\cmidrule{2-5}
& \multicolumn{4}{l}{$F(2, 64) = 9.24$, $p = \textbf{.0003}$, $\eta^2 = 0.22$} \\
\bottomrule
\end{tabular}
\end{table}

\subsection{Signal-to-Noise Ratio}
\label{subsection:SNR}
We performed a comprehensive SNR analysis for each frequency band and each participant to quantify the strength of EEG signals during correct estimation, overestimation, and underestimation states.
An RM-ANOVA was conducted individually for each band to assess statistical significance among overestimation, correct estimation, and underestimation.
The RM-ANOVA results indicate a significant difference between time perception states in all frequency bands except the alpha band ($p<.05$), as shown in~\cref{tab:anova_results}.

Subsequently, we used the post-hoc Tukey-Kramer comparison test to investigate further pairwise differences between the SNR values of the significant bands for each pair of the three time perception states.
The test results between the over- and underestimation states show a statistically significant difference in all bands.
Comparison between the correct estimation and underestimation also reveals significant differences in the delta, theta, and gamma bands.
In the overestimation and correct estimation pair, the difference is significant only in the beta band (cf.~\cref{tab:anova_results}).

\subsection{Time-Frequency Analysis}
\label{subsec:timeFreqAnalysis}
\begin{figure*}[tb]
  \centering
  \includegraphics[width=\linewidth]{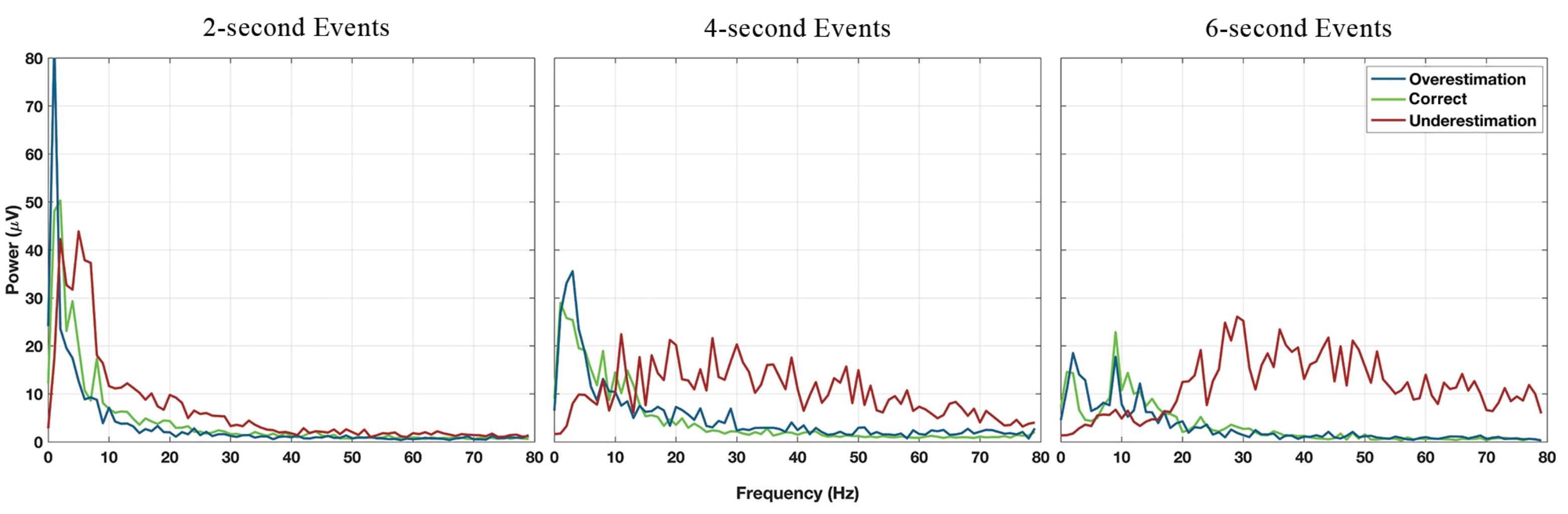}
  \caption{
    Power spectral density averaged across time perception states, grouped by modulation duration.
  }
  \label{fig:psd_duration}
\end{figure*}

Furthermore, we grouped events based on modulation duration and calculated PSD for the three time perception states for each group (\cref{fig:psd_duration}).
While time perception signatures are evident in the PSD plots for 4- and 6-second events, higher-frequency oscillations are absent in the PSD plot for 2-second events.
In general, brain activity in response to a stimulus begins with a transient phase and subsequently shifts into an oscillatory response corresponding to the nature of the stimulus.
The presence of transient responses makes it impossible to capture higher-frequency oscillations.
To validate the presence these responses in 2-second events, we conducted time-frequency analysis on 6-second events, examining the progression of neural dynamics over time.

The results of the analysis for different time perception states are depicted in \cref{fig:teaser}.
All subplots demonstrate that oscillatory responses are observed approximately 1\,second after the onset of the event.
The higher beta band activity is observed during correct estimation (\cref{fig:accurate_estimation}), as well as higher lower-frequency band activities in the overestimation state (\cref{fig:overestimation}).
Similarly, increased high-beta and gamma band activities are evident during underestimation throughout the interval (\cref{fig:underestimation}).

\section{Discussion}
\label{sec:discussion}

Our work provides an important foundation for further research into how time perception, actively used, can complement existing sensory modalities and mental states as an additional component for adaptive VR experiences tailored to the user.
In the following we discuss the key findings as well as limitations and ideas for future work.

\subsection{Time Perception and VR}
\label{subsec:discussion_vr}

The compression effect of VR on time perception was evident in our results, with 57\% of our samples labeled as underestimation.
This effect was not correlated to a specific modulator, demonstrating that it is a general effect.
However, in line with time perception literature, we observed a weak negative correlation between the modulation duration and estimated duration.
While shorter intervals are typically overestimated, longer durations tend to be underestimated~\cite{ grondin2007judging, polti2018effect}.

Except for the emotion-based modulator, all modulators significantly shifted time perception compared to the control events.
However, the different modes of the modulators did not show significant variations in their impact on time perception.
Cui et al.~\cite{cui2023role} comprehensively reviewed time perception studies focusing on emotional valence and arousal over the past 25 years.
Their paper highlights how the effect of emotion on time perception varies according to a range of parameters, including the timing paradigm, the timing task, the secondary task, and the duration of the modulation.
Still, the most frequent observation is the overestimation of emotional stimuli, especially with negative valence~\cite{cui2023role}.
Our results are aligned with this finding, although the observed perception deviation caused by emotion events is not significant.

In contrast, the literature shows better consistency with the effect of cognitive load on time perception.
A higher cognitive load is commonly associated with underestimating duration in prospective timing and overestimating it in retrospective timing~\cite{block2010cognitive, hicks1976prospective, khan2006effect, schatzschneider2016turned}.
Nevertheless, we observed that adding a secondary cognitive task increases perceived duration significantly compared to control events.
Previous work also has reported the same opposite effect~\cite{lallemand2012enhancing,li2020effect}.

The Oddball modulator is unanimously considered to increase perceived duration~\cite{pariyadath2007effect, simchy2018expectation, tachmatzidou2023attention, tse2004attention, ulrich2006perceived}.
We observed the same effect in our study, which was particularly stronger with the contextually relevant mode. 
This finding corroborates previous work of Schweitzer et al.~\cite{schweitzer2017associated}.

For the magnitude modulator, participants significantly overestimated the duration of high-pitch and natural pitch compared to the control condition.
The low-pitch mode showed a significantly different effect on time perception than the natural mode, which was designed as the intra-modulator control condition.
These findings are also supported by the literature investigating the effect of pitch, frequency, and tempo on perceived duration~\cite{beckova2022unraveling, berglund1969influence, cao2019shorten, jeon1997duration, kanai2006time, sunthorn2023making}.

Concerning expectation, studies investigating progress bars show that steady advancement is perceived longer than a non-linear rate.
However, the findings vary depending on whether the decelerating and accelerating progress rates differ in terms of compressing perceived duration~\cite{harrison2007rethinking, kim2017effect, kuroki2015manipulating}.
In our work, while the slow progress bar did not change the perception of duration compared to the control condition, the fast progress bar significantly extended the perceived duration compared to both the slow progress bar and the control condition.

\subsection{EEG Signatures of Time Perception States}
\label{subsec:discussion_eeg_signatures}

Spectral activities revealed distinctive signatures corresponding to each time perception state (\cref{fig:PSD}).
The overestimation samples exhibit a pattern, standing out with lower band activities, prominently in the delta and theta bands.
The correct estimation of intervals appears to be closely associated with an EEG spectral signature characterized by increased activity in the alpha and beta bands, particularly within the low- and medium-beta frequency ranges.
On the other hand, the underestimation state is distinguished by increased higher band powers, specifically in the high-beta and gamma bands.
The topography maps in \cref{fig:topo} further confirm the identifiable patterns specific to time perception states and present a spatial overview of neural dynamics during temporal processing.

The significance of these specific signatures was also quantitatively validated.
The patterns of neural activity proved to be significantly different among the three states and also for each pair of these states.
However, the most striking difference is observed between the over- and underestimation states.
This finding resonates intuitively, as these two states represent the opposite ends of the time perception spectrum.

Time-frequency analysis revealed the presence of transient response in the first half of 2-second events, blocking higher-frequency oscillations (\cref{fig:teaser}).
Most significantly, the activity patterns in the first 2 and 4 seconds of the 6-second events plots in \cref{fig:teaser} demonstrate a remarkable mapping with the PSD of 2- and 4-second events (\cref{fig:psd_duration}), respectively.
This precise mapping ensures that the signatures of time perception states evolve identically, regardless of the event duration.

Our results have been partially observed and confirmed by other studies.
For example, alpha and beta band activities have long been linked to time perception~\cite{ghaderi2018time, minkwitz2012time, mioni2020modulation, wiener2018intrinsic}.
In line with these findings, the increased activity in these two bands is the signature of our samples of correct estimation.
Furthermore, as shown in \cref{fig:topo}, we observed reduced beta activity during overestimation and a contrast between the overestimation and underestimation states in the beta bands.
These observations were also noted by Ghaderi et al.~\cite{ghaderi2018time}.
In another work investigating the effect of task complexity on time estimation in a VR environment, Li and Kim~\cite{li2021effect} reported a relative reduction of the beta band power at the Fz and Pz electrode locations, which was correlated with larger interval estimation errors.
We also observed the same power modulation at Fz and Pz locations for both overestimation and underestimation states compared to the correct estimation (\cref{fig:topo}).
Furthermore, Klimesch~\cite{klimesch2012alpha} noted that a reduction in alpha band power characterizes duration underestimation.
Similarly, we observed minimal activity in the alpha band and substantial modulations in high-beta and gamma band power (\cref{fig:topo}).

\subsection{Support for the Attention Allocation Model}
\label{subsec:discussion_time_theory}

This work focuses on the objective measurement and modulation of time perception in virtual reality rather than explaining its underlying neural mechanisms.
However, our findings reveal cognitive markers that resonate remarkably with the widely recognized attention-based model of time perception, further developed and popularized by Dan Zakay~\cite{zakay1989subjective, zakay1993relative, zakay2014psychological}.
According to this model, our estimation of durations depends on the amount of attention allocated to the passage of time, which is drawn from a shared pool of attentional resources typically prioritized for other, more primary cognitive processes.
For example, in presence of competing attentional demands, we pay less attention to the passage of time and underestimate durations.
Intriguingly, in cases of underestimation, we observed increased neural activity in higher bands, which indicates heightened cognitive engagement or emotional arousal~\cite{yang2020high}.
In other words, participants estimated durations shorter when a mental distraction diverted their attention.
In the case of correct estimation, the signature peak appears in the middle range, around the alpha and lower-beta bands, reflecting a calm yet alert state of mind--capable of perceiving the present and mindful of the passage of time~\cite{hinterberger2014decreased}.
Finally, the overestimation state differs from the correct state by the absence of activity in the beta band, which typically signals a lack of focused attention.
Coupled with increased activity in lower bands, the neural signature of the overestimation state illustrates a drowsy, exhausted, and disengaged cognitive state, leaving all attentional resources available for excessive observation of time~\cite{hinterberger2014decreased}.

\subsection{Limitations and Future Work}
\label{subsec:discussion_limitations}

While we identified distinct signatures of time perception states, our findings are limited to the relatively short time scale of the experiment's events.
Further research is needed to assess the reliability of EEG signatures over longer durations, making them applicable to extended VR experiences.
Moreover, we analyzed the EEG patterns for signatures corresponding to the three discrete time perception states: overestimation, correct estimation, and underestimation.
Moving forward, it will be interesting to investigate features of EEG signals to estimate the magnitude of NEE; that is, to know not only if the user is perceiving time slower or faster, but also how much slower or faster.

Finally, we note that using two separate devices, a VR headset and an EEG cap, was not the most optimal or comfortable option for participants.
Dedicated hardware integrating biosensors with the VR headset, could tackle this issue and improve user experience.

We have measured time perception in VR considering different modes of five modulators in two sensory modalities.
We have established the potential of brain signals as an objective physiological measure of time perception in VR.
We have identified clear spectral signatures for overestimation, correct estimation, and underestimation of time perception states that are persistent across individuals, modulators, and modulation duration.
These signatures can be integrated and applied to monitor and actively influence time perception in VR, allowing virtual environments to be purposefully adapted to the individual, to increase immersion further and improve user experience.

\section*{Supplementary Materials}
\label{sec:supplementary_materials}

A detailed table of arousal and valence values for the selected pictures used in the emotion modulator, animated illustrations of the oddball and expectation modulators, audio files for the magnitude modulator, and a video showcasing an example event are available online\footnote{\url{https://vrarlab.uni.lu/pub/brain-signatures}}.

\acknowledgments{
    This study has been supported by the European Union’s Horizon 2020 research and innovation programme under grant numbers 964464 (ChronoPilot) and CHIST-ERA-20-BCI-001 (BANANA), as well as the European Innovation Council under grant number 101071147 (SYMBIOTIK).
}

\bibliographystyle{abbrv-doi-hyperref-narrow}
\bibliography{BrainSignatures}

\begin{thebibliography}{10}
\renewcommand*{\sfdefault}{PTSansNarrow-TLF}

\bibitem{angrilli1997influence}
\href{https://doi.org/10.3758/bf03205512}{A.~Angrilli, P.~Cherubini, A.~Pavese,
  and S.~Manfredini}.
\newblock \href{https://doi.org/10.3758/bf03205512}{The influence of affective
  factors on time perception}.
\newblock \href{https://doi.org/10.3758/bf03205512}{{\em Perception \&
  Psychophysics}}, \href{https://doi.org/10.3758/bf03205512}{59:972--982},
  \href{https://doi.org/10.3758/bf03205512}{Jan. 1997}.
  \href{https://doi.org/10.3758/bf03205512}
{doi: \textsf{%
10\hspace{.1pt}\discretionary{.}{%
}{.}\hspace{.4pt}3758\discretionary{/}{%
}{/}bf03205512}}


\bibitem{beckova2022unraveling}
\href{https://doi.org/10.1525/mp.2022.40.2.135}{A.~Beckov{\'a},
  V.~Rudolfov{\'a}, J.~Hor{\'a}{\v{c}}ek, and T.~Nekov{\'a}{\v{r}}ov{\'a}}.
\newblock \href{https://doi.org/10.1525/mp.2022.40.2.135}{Unraveling the filled
  duration illusion and its stability in repeated measurements: Acoustic
  stimuli and the possible prospect for time in music}.
\newblock \href{https://doi.org/10.1525/mp.2022.40.2.135}{{\em Music
  Perception: An Interdisciplinary Journal}},
  \href{https://doi.org/10.1525/mp.2022.40.2.135}{40(2):135--149},
  \href{https://doi.org/10.1525/mp.2022.40.2.135}{Jan. 2022}.
  \href{https://doi.org/10.1525/mp.2022.40.2.135}
{doi: \textsf{%
10\hspace{.1pt}\discretionary{.}{%
}{.}\hspace{.4pt}1525\discretionary{/}{%
}{/}mp\hspace{.1pt}\discretionary{.}{%
}{.}\hspace{.4pt}2022\hspace{.1pt}\discretionary{.}{%
}{.}\hspace{.4pt}40\hspace{.1pt}\discretionary{.}{%
}{.}\hspace{.4pt}2\hspace{.1pt}\discretionary{.}{%
}{.}\hspace{.4pt}135}}


\bibitem{berglund1969influence}
\href{https://doi.org/10.1111/j.1467-9450.1969.tb00003.x}{B.~Berglund,
  U.~Berglund, G.~Ekman, and M.~Frankehaeuser}.
\newblock \href{https://doi.org/10.1111/j.1467-9450.1969.tb00003.x}{The
  influence of auditory stimulus intensity on apparent duration}.
\newblock \href{https://doi.org/10.1111/j.1467-9450.1969.tb00003.x}{{\em
  Scandinavian Journal of Psychology}},
  \href{https://doi.org/10.1111/j.1467-9450.1969.tb00003.x}{10(1):21--26},
  \href{https://doi.org/10.1111/j.1467-9450.1969.tb00003.x}{Sept. 1969}.
  \href{https://doi.org/10.1111/j.1467-9450.1969.tb00003.x}
{doi: \textsf{%
10\hspace{.1pt}\discretionary{.}{%
}{.}\hspace{.4pt}1111\discretionary{/}{%
}{/}j\hspace{.1pt}\discretionary{.}{%
}{.}\hspace{.4pt}1467\discretionary{%
}{-}{-}9450\hspace{.1pt}\discretionary{.}{%
}{.}\hspace{.4pt}1969\hspace{.1pt}\discretionary{.}{%
}{.}\hspace{.4pt}tb00003\hspace{.1pt}\discretionary{.}{%
}{.}\hspace{.4pt}x}}


\bibitem{block2010impact}
\href{https://doi.org/10.1145/1868914.1868985}{F.~Block and H.~Gellersen}.
\newblock \href{https://doi.org/10.1145/1868914.1868985}{The impact of
  cognitive load on the perception of time}.
\newblock \href{https://doi.org/10.1145/1868914.1868985}{In {\em Proc.\
  NordiCHI}}, \href{https://doi.org/10.1145/1868914.1868985}{pp. 607--610}.
  \href{https://doi.org/10.1145/1868914.1868985}{ACM},
  \href{https://doi.org/10.1145/1868914.1868985}{New York, NY, USA},
  \href{https://doi.org/10.1145/1868914.1868985}{2010}.
  \href{https://doi.org/10.1145/1868914.1868985}
{doi: \textsf{%
10\hspace{.1pt}\discretionary{.}{%
}{.}\hspace{.4pt}1145\discretionary{/}{%
}{/}1868914\hspace{.1pt}\discretionary{.}{%
}{.}\hspace{.4pt}1868985}}


\bibitem{block2010cognitive}
\href{https://doi.org/10.1016/j.actpsy.2010.03.006}{R.~A. Block, P.~A. Hancock,
  and D.~Zakay}.
\newblock \href{https://doi.org/10.1016/j.actpsy.2010.03.006}{How cognitive
  load affects duration judgments: A meta-analytic review}.
\newblock \href{https://doi.org/10.1016/j.actpsy.2010.03.006}{{\em Acta
  Psychologica}},
  \href{https://doi.org/10.1016/j.actpsy.2010.03.006}{134(3):330--343},
  \href{https://doi.org/10.1016/j.actpsy.2010.03.006}{July 2010}.
  \href{https://doi.org/10.1016/j.actpsy.2010.03.006}
{doi: \textsf{%
10\hspace{.1pt}\discretionary{.}{%
}{.}\hspace{.4pt}1016\discretionary{/}{%
}{/}j\hspace{.1pt}\discretionary{.}{%
}{.}\hspace{.4pt}actpsy\hspace{.1pt}\discretionary{.}{%
}{.}\hspace{.4pt}2010\hspace{.1pt}\discretionary{.}{%
}{.}\hspace{.4pt}03\hspace{.1pt}\discretionary{.}{%
}{.}\hspace{.4pt}006}}


\bibitem{bogon2024age}
\href{https://doi.org/10.1016/j.actpsy.2024.104460}{J.~Bogon, C.~Jagorska,
  I.~Steinecker, and M.~Riemer}.
\newblock \href{https://doi.org/10.1016/j.actpsy.2024.104460}{Age-related
  changes in time perception: Effects of immersive virtual reality and spatial
  location of stimuli}.
\newblock \href{https://doi.org/10.1016/j.actpsy.2024.104460}{{\em Acta
  Psychologica}},
  \href{https://doi.org/10.1016/j.actpsy.2024.104460}{249:104460},
  \href{https://doi.org/10.1016/j.actpsy.2024.104460}{Sept. 2024}.
  \href{https://doi.org/10.1016/j.actpsy.2024.104460}
{doi: \textsf{%
10\hspace{.1pt}\discretionary{.}{%
}{.}\hspace{.4pt}1016\discretionary{/}{%
}{/}j\hspace{.1pt}\discretionary{.}{%
}{.}\hspace{.4pt}actpsy\hspace{.1pt}\discretionary{.}{%
}{.}\hspace{.4pt}2024\hspace{.1pt}\discretionary{.}{%
}{.}\hspace{.4pt}104460}}


\bibitem{botev2021chronopilot}
\href{https://doi.org/10.1109/AIVR52153.2021.00049}{J.~Botev, K.~Drewing,
  H.~Hamann, Y.~Khaluf, P.~Simoens, and A.~Vatakis}.
\newblock \href{https://doi.org/10.1109/AIVR52153.2021.00049}{{C}hrono{P}ilot
  – modulating time perception}.
\newblock \href{https://doi.org/10.1109/AIVR52153.2021.00049}{In {\em Proc.\
  AIVR}}, \href{https://doi.org/10.1109/AIVR52153.2021.00049}{pp. 215--218}.
  \href{https://doi.org/10.1109/AIVR52153.2021.00049}{IEEE},
  \href{https://doi.org/10.1109/AIVR52153.2021.00049}{Piscataway, NJ, USA},
  \href{https://doi.org/10.1109/AIVR52153.2021.00049}{2021}.
  \href{https://doi.org/10.1109/AIVR52153.2021.00049}
{doi: \textsf{%
10\hspace{.1pt}\discretionary{.}{%
}{.}\hspace{.4pt}1109\discretionary{/}{%
}{/}AIVR52153\hspace{.1pt}\discretionary{.}{%
}{.}\hspace{.4pt}2021\hspace{.1pt}\discretionary{.}{%
}{.}\hspace{.4pt}00049}}


\bibitem{bruder2014time}
\href{https://doi.org/10.1109/vr.2014.6802054}{G.~Bruder and F.~Steinicke}.
\newblock \href{https://doi.org/10.1109/vr.2014.6802054}{Time perception during
  walking in virtual environments}.
\newblock \href{https://doi.org/10.1109/vr.2014.6802054}{In {\em Proc.\ IEEE
  VR}}, \href{https://doi.org/10.1109/vr.2014.6802054}{pp. 67--68}.
  \href{https://doi.org/10.1109/vr.2014.6802054}{IEEE},
  \href{https://doi.org/10.1109/vr.2014.6802054}{Piscataway, NJ, USA},
  \href{https://doi.org/10.1109/vr.2014.6802054}{2014}.
  \href{https://doi.org/10.1109/vr.2014.6802054}
{doi: \textsf{%
10\hspace{.1pt}\discretionary{.}{%
}{.}\hspace{.4pt}1109\discretionary{/}{%
}{/}vr\hspace{.1pt}\discretionary{.}{%
}{.}\hspace{.4pt}2014\hspace{.1pt}\discretionary{.}{%
}{.}\hspace{.4pt}6802054}}


\bibitem{cao2019shorten}
\href{https://doi.org/10.1016/j.aap.2018.12.011}{Y.~Cao, X.~Zhuang, and G.~Ma}.
\newblock \href{https://doi.org/10.1016/j.aap.2018.12.011}{Shorten
  pedestrians’ perceived waiting time: The effect of tempo and pitch in
  audible pedestrian signals at red phase}.
\newblock \href{https://doi.org/10.1016/j.aap.2018.12.011}{{\em Accident
  Analysis \& Prevention}},
  \href{https://doi.org/10.1016/j.aap.2018.12.011}{123:336--340},
  \href{https://doi.org/10.1016/j.aap.2018.12.011}{Feb. 2019}.
  \href{https://doi.org/10.1016/j.aap.2018.12.011}
{doi: \textsf{%
10\hspace{.1pt}\discretionary{.}{%
}{.}\hspace{.4pt}1016\discretionary{/}{%
}{/}j\hspace{.1pt}\discretionary{.}{%
}{.}\hspace{.4pt}aap\hspace{.1pt}\discretionary{.}{%
}{.}\hspace{.4pt}2018\hspace{.1pt}\discretionary{.}{%
}{.}\hspace{.4pt}12\hspace{.1pt}\discretionary{.}{%
}{.}\hspace{.4pt}011}}


\bibitem{chirico2016elapsed}
\href{https://doi.org/10.1007/978-3-319-39345-2_65}{A.~Chirico, M.~D’Aiuto,
  M.~Pinto, C.~Milanese, A.~Napoli, F.~Avino, G.~Iodice, G.~Russo,
  M.~De~Laurentiis, G.~Ciliberto, et~al.}
\newblock \href{https://doi.org/10.1007/978-3-319-39345-2_65}{The elapsed time
  during a virtual reality treatment for stressful procedures. a pool analysis
  on breast cancer patients during chemotherapy}.
\newblock \href{https://doi.org/10.1007/978-3-319-39345-2_65}{In {\em Proc.\
  IIMSS}}, \href{https://doi.org/10.1007/978-3-319-39345-2_65}{pp. 731--738}.
  \href{https://doi.org/10.1007/978-3-319-39345-2_65}{Springer},
  \href{https://doi.org/10.1007/978-3-319-39345-2_65}{Cham, Switzerland},
  \href{https://doi.org/10.1007/978-3-319-39345-2_65}{2016}.
  \href{https://doi.org/10.1007/978-3-319-39345-2_65}
{doi: \textsf{%
10\hspace{.1pt}\discretionary{.}{%
}{.}\hspace{.4pt}1007\discretionary{/}{%
}{/}978\discretionary{%
}{-}{-}3\discretionary{%
}{-}{-}319\discretionary{%
}{-}{-}39345\discretionary{%
}{-}{-}2\_65}}


\bibitem{cui2023role}
\href{https://doi.org/10.3758/s13423-022-02148-3}{X.~Cui, Y.~Tian, L.~Zhang,
  Y.~Chen, Y.~Bai, D.~Li, J.~Liu, P.~Gable, and H.~Yin}.
\newblock \href{https://doi.org/10.3758/s13423-022-02148-3}{The role of
  valence, arousal, stimulus type, and temporal paradigm in the efect of
  emotion on time perception: A meta‑analysis}.
\newblock \href{https://doi.org/10.3758/s13423-022-02148-3}{{\em Psychonomic
  Bulletin \& Review}},
  \href{https://doi.org/10.3758/s13423-022-02148-3}{30(1):1--21},
  \href{https://doi.org/10.3758/s13423-022-02148-3}{July 2023}.
  \href{https://doi.org/10.3758/s13423-022-02148-3}
{doi: \textsf{%
10\hspace{.1pt}\discretionary{.}{%
}{.}\hspace{.4pt}3758\discretionary{/}{%
}{/}s13423\discretionary{%
}{-}{-}022\discretionary{%
}{-}{-}02148\discretionary{%
}{-}{-}3}}


\bibitem{dan2011geneva}
\href{https://doi.org/10.3758/s13428-011-0064-1}{E.~S. Dan-Glauser and K.~R.
  Scherer}.
\newblock \href{https://doi.org/10.3758/s13428-011-0064-1}{The {G}eneva
  affective picture database ({GAPED}): a new 730-picture database focusing on
  valence and normative significance}.
\newblock \href{https://doi.org/10.3758/s13428-011-0064-1}{{\em Behavior
  research methods}},
  \href{https://doi.org/10.3758/s13428-011-0064-1}{43:468--477},
  \href{https://doi.org/10.3758/s13428-011-0064-1}{Mar. 2011}.
  \href{https://doi.org/10.3758/s13428-011-0064-1}
{doi: \textsf{%
10\hspace{.1pt}\discretionary{.}{%
}{.}\hspace{.4pt}3758\discretionary{/}{%
}{/}s13428\discretionary{%
}{-}{-}011\discretionary{%
}{-}{-}0064\discretionary{%
}{-}{-}1}}


\bibitem{de2023altering}
\href{https://doi.org/10.1109/toh.2023.3270639}{Y.~De-Pra, V.~Catrambone,
  V.~van Wassenhove, A.~Moscatelli, G.~Valenza, and M.~Bianchi}.
\newblock \href{https://doi.org/10.1109/toh.2023.3270639}{Altering time
  perception in virtual reality through multimodal visual-tactile kappa
  effect}.
\newblock \href{https://doi.org/10.1109/toh.2023.3270639}{{\em IEEE
  Transactions on Haptics}},
  \href{https://doi.org/10.1109/toh.2023.3270639}{16(4):518--523},
  \href{https://doi.org/10.1109/toh.2023.3270639}{Oct. {--} Dec. 2023}.
  \href{https://doi.org/10.1109/toh.2023.3270639}
{doi: \textsf{%
10\hspace{.1pt}\discretionary{.}{%
}{.}\hspace{.4pt}1109\discretionary{/}{%
}{/}toh\hspace{.1pt}\discretionary{.}{%
}{.}\hspace{.4pt}2023\hspace{.1pt}\discretionary{.}{%
}{.}\hspace{.4pt}3270639}}


\bibitem{delorme2004eeglab}
\href{https://doi.org/10.1016/j.jneumeth.2003.10.009}{A.~Delorme and
  S.~Makeig}.
\newblock \href{https://doi.org/10.1016/j.jneumeth.2003.10.009}{{EEGLAB}: an
  open source toolbox for analysis of single-trial {EEG} dynamics including
  independent component analysis}.
\newblock \href{https://doi.org/10.1016/j.jneumeth.2003.10.009}{{\em Journal of
  Neuroscience Methods}},
  \href{https://doi.org/10.1016/j.jneumeth.2003.10.009}{134(1):9--21},
  \href{https://doi.org/10.1016/j.jneumeth.2003.10.009}{Mar. 2004}.
  \href{https://doi.org/10.1016/j.jneumeth.2003.10.009}
{doi: \textsf{%
10\hspace{.1pt}\discretionary{.}{%
}{.}\hspace{.4pt}1016\discretionary{/}{%
}{/}j\hspace{.1pt}\discretionary{.}{%
}{.}\hspace{.4pt}jneumeth\hspace{.1pt}\discretionary{.}{%
}{.}\hspace{.4pt}2003\hspace{.1pt}\discretionary{.}{%
}{.}\hspace{.4pt}10\hspace{.1pt}\discretionary{.}{%
}{.}\hspace{.4pt}009}}


\bibitem{droit2010time}
\href{https://doi.org/10.1016/j.actpsy.2010.07.003}{S.~Droit-Volet, E.~Bigand,
  D.~Ramos, and J.~L.~O. Bueno}.
\newblock \href{https://doi.org/10.1016/j.actpsy.2010.07.003}{Time flies with
  music whatever its emotional valence}.
\newblock \href{https://doi.org/10.1016/j.actpsy.2010.07.003}{{\em Acta
  Psychologica}},
  \href{https://doi.org/10.1016/j.actpsy.2010.07.003}{135(2):226--232},
  \href{https://doi.org/10.1016/j.actpsy.2010.07.003}{Oct. 2010}.
  \href{https://doi.org/10.1016/j.actpsy.2010.07.003}
{doi: \textsf{%
10\hspace{.1pt}\discretionary{.}{%
}{.}\hspace{.4pt}1016\discretionary{/}{%
}{/}j\hspace{.1pt}\discretionary{.}{%
}{.}\hspace{.4pt}actpsy\hspace{.1pt}\discretionary{.}{%
}{.}\hspace{.4pt}2010\hspace{.1pt}\discretionary{.}{%
}{.}\hspace{.4pt}07\hspace{.1pt}\discretionary{.}{%
}{.}\hspace{.4pt}003}}


\bibitem{droit2007emotions}
\href{https://doi.org/10.1016/j.tics.2007.09.008}{S.~Droit-Volet and W.~H.
  Meck}.
\newblock \href{https://doi.org/10.1016/j.tics.2007.09.008}{How emotions colour
  our perception of time}.
\newblock \href{https://doi.org/10.1016/j.tics.2007.09.008}{{\em Trends in
  Cognitive Sciences}},
  \href{https://doi.org/10.1016/j.tics.2007.09.008}{11(12):504--513},
  \href{https://doi.org/10.1016/j.tics.2007.09.008}{Dec. 2007}.
  \href{https://doi.org/10.1016/j.tics.2007.09.008}
{doi: \textsf{%
10\hspace{.1pt}\discretionary{.}{%
}{.}\hspace{.4pt}1016\discretionary{/}{%
}{/}j\hspace{.1pt}\discretionary{.}{%
}{.}\hspace{.4pt}tics\hspace{.1pt}\discretionary{.}{%
}{.}\hspace{.4pt}2007\hspace{.1pt}\discretionary{.}{%
}{.}\hspace{.4pt}09\hspace{.1pt}\discretionary{.}{%
}{.}\hspace{.4pt}008}}


\bibitem{ernst2017p3}
\href{https://doi.org/10.1016/j.neuropsychologia.2017.06.033}{B.~Ernst, S.~M.
  Reichard, R.~F. Riepl, R.~Steinhauser, S.~F. Zimmermann, and M.~Steinhauser}.
\newblock \href{https://doi.org/10.1016/j.neuropsychologia.2017.06.033}{The
  {P3} and the subjective experience of time}.
\newblock \href{https://doi.org/10.1016/j.neuropsychologia.2017.06.033}{{\em
  Neuropsychologia}},
  \href{https://doi.org/10.1016/j.neuropsychologia.2017.06.033}{103:12--19},
  \href{https://doi.org/10.1016/j.neuropsychologia.2017.06.033}{Aug. 2017}.
  \href{https://doi.org/10.1016/j.neuropsychologia.2017.06.033}
{doi: \textsf{%
10\hspace{.1pt}\discretionary{.}{%
}{.}\hspace{.4pt}1016\discretionary{/}{%
}{/}j\hspace{.1pt}\discretionary{.}{%
}{.}\hspace{.4pt}neuropsychologia\hspace{.1pt}\discretionary{.}{%
}{.}\hspace{.4pt}2017\hspace{.1pt}\discretionary{.}{%
}{.}\hspace{.4pt}06\hspace{.1pt}\discretionary{.}{%
}{.}\hspace{.4pt}033}}


\bibitem{fabbri2012theory}
\href{https://doi.org/10.1016/j.actpsy.2011.09.006}{M.~Fabbri, J.~Cancellieri,
  and V.~Natale}.
\newblock \href{https://doi.org/10.1016/j.actpsy.2011.09.006}{The {A} {T}heory
  {O}f {M}agnitude ({ATOM}) model in temporal perception and reproduction
  tasks}.
\newblock \href{https://doi.org/10.1016/j.actpsy.2011.09.006}{{\em Acta
  Psychologica}},
  \href{https://doi.org/10.1016/j.actpsy.2011.09.006}{139(1):111--123},
  \href{https://doi.org/10.1016/j.actpsy.2011.09.006}{Jan. 2012}.
  \href{https://doi.org/10.1016/j.actpsy.2011.09.006}
{doi: \textsf{%
10\hspace{.1pt}\discretionary{.}{%
}{.}\hspace{.4pt}1016\discretionary{/}{%
}{/}j\hspace{.1pt}\discretionary{.}{%
}{.}\hspace{.4pt}actpsy\hspace{.1pt}\discretionary{.}{%
}{.}\hspace{.4pt}2011\hspace{.1pt}\discretionary{.}{%
}{.}\hspace{.4pt}09\hspace{.1pt}\discretionary{.}{%
}{.}\hspace{.4pt}006}}


\bibitem{faul2007g}
\href{https://doi.org/10.3758/bf03193146}{F.~Faul, E.~Erdfelder, A.-G. Lang,
  and A.~Buchner}.
\newblock \href{https://doi.org/10.3758/bf03193146}{{G*P}ower 3: A flexible
  statistical power analysis program for the social, behavioral, and biomedical
  sciences}.
\newblock \href{https://doi.org/10.3758/bf03193146}{{\em Behavior Research
  Methods}}, \href{https://doi.org/10.3758/bf03193146}{39(2):175--191},
  \href{https://doi.org/10.3758/bf03193146}{May 2007}.
  \href{https://doi.org/10.3758/bf03193146}
{doi: \textsf{%
10\hspace{.1pt}\discretionary{.}{%
}{.}\hspace{.4pt}3758\discretionary{/}{%
}{/}bf03193146}}


\bibitem{fischer2022time}
\href{https://hasisaurus.at/publications/2022_ewsn_Time.pdf}{S.~Fischer,
  L.~Breitsameter, and G.~H{\"o}lzl}.
\newblock \href{https://hasisaurus.at/publications/2022_ewsn_Time.pdf}{Time
  anomalies in virtual reality-impact of manipulated zeitgebers on individual
  human time perception}.
\newblock \href{https://hasisaurus.at/publications/2022_ewsn_Time.pdf}{In {\em
  Proc.\ EWSN}},
  \href{https://hasisaurus.at/publications/2022_ewsn_Time.pdf}{pp. 232--237}.
  \href{https://hasisaurus.at/publications/2022_ewsn_Time.pdf}{ACM},
  \href{https://hasisaurus.at/publications/2022_ewsn_Time.pdf}{New York, NY,
  USA}, \href{https://hasisaurus.at/publications/2022_ewsn_Time.pdf}{2022}.

\bibitem{ghaderi2018time}
\href{https://doi.org/10.1371/journal.pone.0195380}{A.~H. Ghaderi,
  S.~Moradkhani, A.~Haghighatfard, F.~Akrami, Z.~Khayyer, and F.~Balc{\i}}.
\newblock \href{https://doi.org/10.1371/journal.pone.0195380}{Time estimation
  and beta segregation: An {EEG} study and graph theoretical approach}.
\newblock \href{https://doi.org/10.1371/journal.pone.0195380}{{\em PLoS One}},
  \href{https://doi.org/10.1371/journal.pone.0195380}{13(4):e0195380},
  \href{https://doi.org/10.1371/journal.pone.0195380}{Apr. 2018}.
  \href{https://doi.org/10.1371/journal.pone.0195380}
{doi: \textsf{%
10\hspace{.1pt}\discretionary{.}{%
}{.}\hspace{.4pt}1371\discretionary{/}{%
}{/}journal\hspace{.1pt}\discretionary{.}{%
}{.}\hspace{.4pt}pone\hspace{.1pt}\discretionary{.}{%
}{.}\hspace{.4pt}0195380}}


\bibitem{grabot2019strength}
\href{https://doi.org/10.1523/JNEUROSCI.2473-18.2018}{L.~Grabot, T.~W.
  Kononowicz, T.~D. La~Tour, A.~Gramfort, V.~Doy{\`e}re, and V.~van
  Wassenhove}.
\newblock \href{https://doi.org/10.1523/JNEUROSCI.2473-18.2018}{The strength of
  alpha–beta oscillatory coupling predicts motor timing precision}.
\newblock \href{https://doi.org/10.1523/JNEUROSCI.2473-18.2018}{{\em Journal of
  Neuroscience}},
  \href{https://doi.org/10.1523/JNEUROSCI.2473-18.2018}{39(17):3277--3291},
  \href{https://doi.org/10.1523/JNEUROSCI.2473-18.2018}{Apr. 2019}.
  \href{https://doi.org/10.1523/JNEUROSCI.2473-18.2018}
{doi: \textsf{%
10\hspace{.1pt}\discretionary{.}{%
}{.}\hspace{.4pt}1523\discretionary{/}{%
}{/}JNEUROSCI\hspace{.1pt}\discretionary{.}{%
}{.}\hspace{.4pt}2473\discretionary{%
}{-}{-}18\hspace{.1pt}\discretionary{.}{%
}{.}\hspace{.4pt}2018}}


\bibitem{grondin2007judging}
\href{https://doi.org/10.1080/17470210600988976}{S.~Grondin and M.~Plourde}.
\newblock \href{https://doi.org/10.1080/17470210600988976}{Judging multi-minute
  intervals retrospectively}.
\newblock \href{https://doi.org/10.1080/17470210600988976}{{\em Quarterly
  Journal of Experimental Psychology}},
  \href{https://doi.org/10.1080/17470210600988976}{60(9):1303--1312},
  \href{https://doi.org/10.1080/17470210600988976}{Aug. 2007}.
  \href{https://doi.org/10.1080/17470210600988976}
{doi: \textsf{%
10\hspace{.1pt}\discretionary{.}{%
}{.}\hspace{.4pt}1080\discretionary{/}{%
}{/}17470210600988976}}


\bibitem{harrison2007rethinking}
\href{https://doi.org/10.1145/1294211.1294231}{C.~Harrison, B.~Amento,
  S.~Kuznetsov, and R.~Bell}.
\newblock \href{https://doi.org/10.1145/1294211.1294231}{Rethinking the
  progress bar}.
\newblock \href{https://doi.org/10.1145/1294211.1294231}{In {\em Proc.\ UIST}},
  \href{https://doi.org/10.1145/1294211.1294231}{pp. 115--118}.
  \href{https://doi.org/10.1145/1294211.1294231}{ACM},
  \href{https://doi.org/10.1145/1294211.1294231}{New York, NY, USA},
  \href{https://doi.org/10.1145/1294211.1294231}{2007}.
  \href{https://doi.org/10.1145/1294211.1294231}
{doi: \textsf{%
10\hspace{.1pt}\discretionary{.}{%
}{.}\hspace{.4pt}1145\discretionary{/}{%
}{/}1294211\hspace{.1pt}\discretionary{.}{%
}{.}\hspace{.4pt}1294231}}


\bibitem{hicks1976prospective}
\href{https://doi.org/10.2307/1421469}{R.~E. Hicks, G.~W. Miller, and
  M.~Kinsbourne}.
\newblock \href{https://doi.org/10.2307/1421469}{Prospective and retrospective
  judgments of time as a function of amount of information processed}.
\newblock \href{https://doi.org/10.2307/1421469}{{\em The American Journal of
  Psychology}}, \href{https://doi.org/10.2307/1421469}{89(4):719--730},
  \href{https://doi.org/10.2307/1421469}{Dec. 1976}.
  \href{https://doi.org/10.2307/1421469}
{doi: \textsf{%
10\hspace{.1pt}\discretionary{.}{%
}{.}\hspace{.4pt}2307\discretionary{/}{%
}{/}1421469}}


\bibitem{hinterberger2014decreased}
\href{https://doi.org/10.3389/fpsyg.2014.00099}{T.~Hinterberger, S.~Schmidt,
  T.~Kamei, and H.~Walach}.
\newblock \href{https://doi.org/10.3389/fpsyg.2014.00099}{Decreased
  electrophysiological activity represents the conscious state of emptiness in
  meditation}.
\newblock \href{https://doi.org/10.3389/fpsyg.2014.00099}{{\em Frontiers in
  psychology}}, \href{https://doi.org/10.3389/fpsyg.2014.00099}{5:99},
  \href{https://doi.org/10.3389/fpsyg.2014.00099}{Feb. 2014}.
  \href{https://doi.org/10.3389/fpsyg.2014.00099}
{doi: \textsf{%
10\hspace{.1pt}\discretionary{.}{%
}{.}\hspace{.4pt}3389\discretionary{/}{%
}{/}fpsyg\hspace{.1pt}\discretionary{.}{%
}{.}\hspace{.4pt}2014\hspace{.1pt}\discretionary{.}{%
}{.}\hspace{.4pt}00099}}


\bibitem{horr2016perceived}
\href{https://doi.org/10.1016/j.neuroimage.2016.02.011}{N.~K. Horr, M.~Wimber,
  and M.~Di~Luca}.
\newblock \href{https://doi.org/10.1016/j.neuroimage.2016.02.011}{Perceived
  time and temporal structure: Neural entrainment to isochronous stimulation
  increases duration estimates}.
\newblock \href{https://doi.org/10.1016/j.neuroimage.2016.02.011}{{\em
  Neuroimage}},
  \href{https://doi.org/10.1016/j.neuroimage.2016.02.011}{132:148--156},
  \href{https://doi.org/10.1016/j.neuroimage.2016.02.011}{May 2016}.
  \href{https://doi.org/10.1016/j.neuroimage.2016.02.011}
{doi: \textsf{%
10\hspace{.1pt}\discretionary{.}{%
}{.}\hspace{.4pt}1016\discretionary{/}{%
}{/}j\hspace{.1pt}\discretionary{.}{%
}{.}\hspace{.4pt}neuroimage\hspace{.1pt}\discretionary{.}{%
}{.}\hspace{.4pt}2016\hspace{.1pt}\discretionary{.}{%
}{.}\hspace{.4pt}02\hspace{.1pt}\discretionary{.}{%
}{.}\hspace{.4pt}011}}


\bibitem{igarzabal2021happens}
\href{https://doi.org/10.1037/tmb0000038}{F.~A. Igarz{\'a}bal, H.~Hruby,
  J.~Witowska, S.~Khoshnoud, and M.~Wittmann}.
\newblock \href{https://doi.org/10.1037/tmb0000038}{What happens while waiting
  in virtual reality? a comparison between a virtual and a real waiting
  situation concerning boredom, self-regulation, and the experience of time}.
\newblock \href{https://doi.org/10.1037/tmb0000038}{{\em Technology, Mind, and
  Behavior}}, \href{https://doi.org/10.1037/tmb0000038}{2(2)},
  \href{https://doi.org/10.1037/tmb0000038}{July 2021}.
  \href{https://doi.org/10.1037/tmb0000038}
{doi: \textsf{%
10\hspace{.1pt}\discretionary{.}{%
}{.}\hspace{.4pt}1037\discretionary{/}{%
}{/}tmb0000038}}


\bibitem{jeon1997duration}
\href{https://doi.org/10.1177/0305735697251006}{J.~Y. Jeon and F.~R. Fricke}.
\newblock \href{https://doi.org/10.1177/0305735697251006}{Duration of perceived
  and performed sounds}.
\newblock \href{https://doi.org/10.1177/0305735697251006}{{\em Psychology of
  Music}}, \href{https://doi.org/10.1177/0305735697251006}{25(1):70--83},
  \href{https://doi.org/10.1177/0305735697251006}{Apr. 1997}.
  \href{https://doi.org/10.1177/0305735697251006}
{doi: \textsf{%
10\hspace{.1pt}\discretionary{.}{%
}{.}\hspace{.4pt}1177\discretionary{/}{%
}{/}0305735697251006}}


\bibitem{kanai2006time}
\href{https://doi.org/10.1167/6.12.8}{R.~Kanai, C.~L. Paffen, H.~Hogendoorn,
  and F.~A. Verstraten}.
\newblock \href{https://doi.org/10.1167/6.12.8}{Time dilation in dynamic visual
  display}.
\newblock \href{https://doi.org/10.1167/6.12.8}{{\em Journal of Vision}},
  \href{https://doi.org/10.1167/6.12.8}{6(12):8--8},
  \href{https://doi.org/10.1167/6.12.8}{Dec. 2006}.
  \href{https://doi.org/10.1167/6.12.8}
{doi: \textsf{%
10\hspace{.1pt}\discretionary{.}{%
}{.}\hspace{.4pt}1167\discretionary{/}{%
}{/}6\hspace{.1pt}\discretionary{.}{%
}{.}\hspace{.4pt}12\hspace{.1pt}\discretionary{.}{%
}{.}\hspace{.4pt}8}}


\bibitem{kaneko2009perceived}
\href{https://doi.org/10.1167/9.7.14}{S.~Kaneko and I.~Murakami}.
\newblock \href{https://doi.org/10.1167/9.7.14}{Perceived duration of visual
  motion increases with speed}.
\newblock \href{https://doi.org/10.1167/9.7.14}{{\em Journal of Vision}},
  \href{https://doi.org/10.1167/9.7.14}{9(7):14--14},
  \href{https://doi.org/10.1167/9.7.14}{July 2009}.
  \href{https://doi.org/10.1167/9.7.14}
{doi: \textsf{%
10\hspace{.1pt}\discretionary{.}{%
}{.}\hspace{.4pt}1167\discretionary{/}{%
}{/}9\hspace{.1pt}\discretionary{.}{%
}{.}\hspace{.4pt}7\hspace{.1pt}\discretionary{.}{%
}{.}\hspace{.4pt}14}}


\bibitem{khan2006effect}
\href{https://www.researchgate.net/publication/235954148_Effect_of_Cognitive_Load_and_Paradigm_on_Time_Perception}{A.~Khan,
  N.~K. Sharma, and S.~Dixit}.
\newblock
  \href{https://www.researchgate.net/publication/235954148_Effect_of_Cognitive_Load_and_Paradigm_on_Time_Perception}{Effect
  of cognitive load and paradigm on time perception}.
\newblock
  \href{https://www.researchgate.net/publication/235954148_Effect_of_Cognitive_Load_and_Paradigm_on_Time_Perception}{{\em
  Journal of the Indian Academy of Applied Psychology}},
  \href{https://www.researchgate.net/publication/235954148_Effect_of_Cognitive_Load_and_Paradigm_on_Time_Perception}{32(1):37--42},
  \href{https://www.researchgate.net/publication/235954148_Effect_of_Cognitive_Load_and_Paradigm_on_Time_Perception}{Jan.
  2006}.

\bibitem{kim2017effect}
\href{https://doi.org/10.1007/978-3-319-58640-3_9}{W.~Kim and S.~Xiong}.
\newblock \href{https://doi.org/10.1007/978-3-319-58640-3_9}{The effect of
  video loading symbol on waiting time perception}.
\newblock \href{https://doi.org/10.1007/978-3-319-58640-3_9}{In {\em Proc.\
  DUXU}}, \href{https://doi.org/10.1007/978-3-319-58640-3_9}{pp. 105--114}.
  \href{https://doi.org/10.1007/978-3-319-58640-3_9}{Springer},
  \href{https://doi.org/10.1007/978-3-319-58640-3_9}{Cham, Switzerland},
  \href{https://doi.org/10.1007/978-3-319-58640-3_9}{2017}.
  \href{https://doi.org/10.1007/978-3-319-58640-3_9}
{doi: \textsf{%
10\hspace{.1pt}\discretionary{.}{%
}{.}\hspace{.4pt}1007\discretionary{/}{%
}{/}978\discretionary{%
}{-}{-}3\discretionary{%
}{-}{-}319\discretionary{%
}{-}{-}58640\discretionary{%
}{-}{-}3\_9}}


\bibitem{klimesch2012alpha}
\href{https://doi.org/10.1016/j.tics.2012.10.007}{W.~Klimesch}.
\newblock \href{https://doi.org/10.1016/j.tics.2012.10.007}{Alpha-band
  oscillations, attention, and controlled access to stored information}.
\newblock \href{https://doi.org/10.1016/j.tics.2012.10.007}{{\em Trends in
  Cognitive Sciences}},
  \href{https://doi.org/10.1016/j.tics.2012.10.007}{16(12):606--617},
  \href{https://doi.org/10.1016/j.tics.2012.10.007}{Dec. 2012}.
  \href{https://doi.org/10.1016/j.tics.2012.10.007}
{doi: \textsf{%
10\hspace{.1pt}\discretionary{.}{%
}{.}\hspace{.4pt}1016\discretionary{/}{%
}{/}j\hspace{.1pt}\discretionary{.}{%
}{.}\hspace{.4pt}tics\hspace{.1pt}\discretionary{.}{%
}{.}\hspace{.4pt}2012\hspace{.1pt}\discretionary{.}{%
}{.}\hspace{.4pt}10\hspace{.1pt}\discretionary{.}{%
}{.}\hspace{.4pt}007}}


\bibitem{kuroki2015manipulating}
\href{https://doi.org/10.1007/978-3-319-21383-5_113}{Y.~Kuroki and
  M.~Ishihara}.
\newblock \href{https://doi.org/10.1007/978-3-319-21383-5_113}{Manipulating
  animation speed of progress bars to shorten time perception}.
\newblock \href{https://doi.org/10.1007/978-3-319-21383-5_113}{In {\em Proc.\
  HCI International}}, \href{https://doi.org/10.1007/978-3-319-21383-5_113}{pp.
  670--673}. \href{https://doi.org/10.1007/978-3-319-21383-5_113}{Springer},
  \href{https://doi.org/10.1007/978-3-319-21383-5_113}{Cham, Switzerland},
  \href{https://doi.org/10.1007/978-3-319-21383-5_113}{2015}.
  \href{https://doi.org/10.1007/978-3-319-21383-5_113}
{doi: \textsf{%
10\hspace{.1pt}\discretionary{.}{%
}{.}\hspace{.4pt}1007\discretionary{/}{%
}{/}978\discretionary{%
}{-}{-}3\discretionary{%
}{-}{-}319\discretionary{%
}{-}{-}21383\discretionary{%
}{-}{-}5\_113}}


\bibitem{lallemand2012enhancing}
\href{https://doi.org/10.1145/2317956.2318069}{C.~Lallemand and G.~Gronier}.
\newblock \href{https://doi.org/10.1145/2317956.2318069}{Enhancing user
  experience during waiting time in {HCI}: contributions of cognitive
  psychology}.
\newblock \href{https://doi.org/10.1145/2317956.2318069}{In {\em Proc.\ DIS}},
  \href{https://doi.org/10.1145/2317956.2318069}{pp. 751--760}.
  \href{https://doi.org/10.1145/2317956.2318069}{ACM},
  \href{https://doi.org/10.1145/2317956.2318069}{New York, NY, USA},
  \href{https://doi.org/10.1145/2317956.2318069}{2012}.
  \href{https://doi.org/10.1145/2317956.2318069}
{doi: \textsf{%
10\hspace{.1pt}\discretionary{.}{%
}{.}\hspace{.4pt}1145\discretionary{/}{%
}{/}2317956\hspace{.1pt}\discretionary{.}{%
}{.}\hspace{.4pt}2318069}}


\bibitem{li2020effect}
\href{https://digital.lib.washington.edu/server/api/core/bitstreams/7f9820e1-80e0-4765-bc58-3c43795f8db2/content}{J.~Li}.
\newblock
  \href{https://digital.lib.washington.edu/server/api/core/bitstreams/7f9820e1-80e0-4765-bc58-3c43795f8db2/content}{{\em
  The Effect of Task Complexity on Time Perception in the Virtual Reality
  Environment: An {EEG} Study}}.
\newblock
  \href{https://digital.lib.washington.edu/server/api/core/bitstreams/7f9820e1-80e0-4765-bc58-3c43795f8db2/content}{University
  of Washington},
  \href{https://digital.lib.washington.edu/server/api/core/bitstreams/7f9820e1-80e0-4765-bc58-3c43795f8db2/content}{Seattle,
  WA, USA},
  \href{https://digital.lib.washington.edu/server/api/core/bitstreams/7f9820e1-80e0-4765-bc58-3c43795f8db2/content}{2020}.

\bibitem{li2021effect}
\href{https://doi.org/10.3390/app11209779}{J.~Li and J.-E. Kim}.
\newblock \href{https://doi.org/10.3390/app11209779}{The effect of task
  complexity on time estimation in the virtual reality environment: An {EEG}
  study}.
\newblock \href{https://doi.org/10.3390/app11209779}{{\em Applied Sciences}},
  \href{https://doi.org/10.3390/app11209779}{11(20):9779},
  \href{https://doi.org/10.3390/app11209779}{Oct. 2021}.
  \href{https://doi.org/10.3390/app11209779}
{doi: \textsf{%
10\hspace{.1pt}\discretionary{.}{%
}{.}\hspace{.4pt}3390\discretionary{/}{%
}{/}app11209779}}


\bibitem{lu2011weight}
\href{https://doi.org/10.1037/a0024673}{A.~Lu, L.~Mo, and B.~H. Hodges}.
\newblock \href{https://doi.org/10.1037/a0024673}{The weight of time:
  Affordances for an integrated magnitude system}.
\newblock \href{https://doi.org/10.1037/a0024673}{{\em Journal of Experimental
  Psychology: Human Perception and Performance}},
  \href{https://doi.org/10.1037/a0024673}{37(6):1855},
  \href{https://doi.org/10.1037/a0024673}{July 2011}.
  \href{https://doi.org/10.1037/a0024673}
{doi: \textsf{%
10\hspace{.1pt}\discretionary{.}{%
}{.}\hspace{.4pt}1037\discretionary{/}{%
}{/}a0024673}}


\bibitem{lugrin2019experiencing}
\href{https://doi.org/10.1145/3359996.3364807}{J.-L. Lugrin, F.~Unruh,
  M.~Landeck, Y.~Lamour, M.~E. Latoschik, K.~Vogeley, and M.~Wittmann}.
\newblock \href{https://doi.org/10.1145/3359996.3364807}{Experiencing waiting
  time in virtual reality}.
\newblock \href{https://doi.org/10.1145/3359996.3364807}{In {\em Proc.\ VRST}},
  \href{https://doi.org/10.1145/3359996.3364807}{pp. 1--2}.
  \href{https://doi.org/10.1145/3359996.3364807}{ACM},
  \href{https://doi.org/10.1145/3359996.3364807}{New York, NY, USA},
  \href{https://doi.org/10.1145/3359996.3364807}{2019}.
  \href{https://doi.org/10.1145/3359996.3364807}
{doi: \textsf{%
10\hspace{.1pt}\discretionary{.}{%
}{.}\hspace{.4pt}1145\discretionary{/}{%
}{/}3359996\hspace{.1pt}\discretionary{.}{%
}{.}\hspace{.4pt}3364807}}


\bibitem{malpica2022larger}
\href{https://doi.org/10.1371/journal.pone.0265591}{S.~Malpica, B.~Masia,
  L.~Herman, G.~Wetzstein, D.~M. Eagleman, D.~Gutierrez, Z.~Bylinskii, and
  Q.~Sun}.
\newblock \href{https://doi.org/10.1371/journal.pone.0265591}{Larger visual
  changes compress time: The inverted effect of asemantic visual features on
  interval time perception}.
\newblock \href{https://doi.org/10.1371/journal.pone.0265591}{{\em PLoS ONE}},
  \href{https://doi.org/10.1371/journal.pone.0265591}{17(3):e0265591},
  \href{https://doi.org/10.1371/journal.pone.0265591}{Mar. 2022}.
  \href{https://doi.org/10.1371/journal.pone.0265591}
{doi: \textsf{%
10\hspace{.1pt}\discretionary{.}{%
}{.}\hspace{.4pt}1371\discretionary{/}{%
}{/}journal\hspace{.1pt}\discretionary{.}{%
}{.}\hspace{.4pt}pone\hspace{.1pt}\discretionary{.}{%
}{.}\hspace{.4pt}0265591}}


\bibitem{marchewka2014nencki}
\href{https://doi.org/10.3758/s13428-013-0379-1}{A.~Marchewka,
  {\L}.~{\.Z}urawski, K.~Jednor{\'o}g, and A.~Grabowska}.
\newblock \href{https://doi.org/10.3758/s13428-013-0379-1}{The {N}encki
  {A}ffective {P}icture {S}ystem ({NAPS}): Introduction to a novel,
  standardized, wide-range, high-quality, realistic picture database}.
\newblock \href{https://doi.org/10.3758/s13428-013-0379-1}{{\em Behavior
  Research Methods}},
  \href{https://doi.org/10.3758/s13428-013-0379-1}{46:596--610},
  \href{https://doi.org/10.3758/s13428-013-0379-1}{Aug. 2014}.
  \href{https://doi.org/10.3758/s13428-013-0379-1}
{doi: \textsf{%
10\hspace{.1pt}\discretionary{.}{%
}{.}\hspace{.4pt}3758\discretionary{/}{%
}{/}s13428\discretionary{%
}{-}{-}013\discretionary{%
}{-}{-}0379\discretionary{%
}{-}{-}1}}


\bibitem{martins2022non}
\href{https://doi.org/10.1080/00207454.2020.1826945}{D.~C. Martins~e Silva,
  V.~Marinho, S.~Teixeira, G.~Teles, J.~Marques, A.~Esc{\'o}rcio, T.~Fernandes,
  A.~C. Freitas, M.~Nunes, M.~Ayres, et~al.}
\newblock \href{https://doi.org/10.1080/00207454.2020.1826945}{Non-immersive
  {3D} virtual stimulus alter the time production task performance and increase
  the {EEG} theta power in dorsolateral prefrontal cortex}.
\newblock \href{https://doi.org/10.1080/00207454.2020.1826945}{{\em
  International Journal of Neuroscience}},
  \href{https://doi.org/10.1080/00207454.2020.1826945}{132(6):563--573},
  \href{https://doi.org/10.1080/00207454.2020.1826945}{Oct. 2022}.
  \href{https://doi.org/10.1080/00207454.2020.1826945}
{doi: \textsf{%
10\hspace{.1pt}\discretionary{.}{%
}{.}\hspace{.4pt}1080\discretionary{/}{%
}{/}00207454\hspace{.1pt}\discretionary{.}{%
}{.}\hspace{.4pt}2020\hspace{.1pt}\discretionary{.}{%
}{.}\hspace{.4pt}1826945}}


\bibitem{matthews2016repetition}
\href{https://doi.org/10.1016/j.cobeha.2016.02.019}{W.~J. Matthews and A.~I.
  Gheorghiu}.
\newblock \href{https://doi.org/10.1016/j.cobeha.2016.02.019}{Repetition,
  expectation, and the perception of time}.
\newblock \href{https://doi.org/10.1016/j.cobeha.2016.02.019}{{\em Current
  Opinion in Behavioral Sciences}},
  \href{https://doi.org/10.1016/j.cobeha.2016.02.019}{8:110--116},
  \href{https://doi.org/10.1016/j.cobeha.2016.02.019}{Apr. 2016}.
  \href{https://doi.org/10.1016/j.cobeha.2016.02.019}
{doi: \textsf{%
10\hspace{.1pt}\discretionary{.}{%
}{.}\hspace{.4pt}1016\discretionary{/}{%
}{/}j\hspace{.1pt}\discretionary{.}{%
}{.}\hspace{.4pt}cobeha\hspace{.1pt}\discretionary{.}{%
}{.}\hspace{.4pt}2016\hspace{.1pt}\discretionary{.}{%
}{.}\hspace{.4pt}02\hspace{.1pt}\discretionary{.}{%
}{.}\hspace{.4pt}019}}


\bibitem{minkwitz2012time}
\href{https://doi.org/10.1186/1744-9081-8-50}{J.~Minkwitz, M.~U. Trenner,
  C.~Sander, S.~Olbrich, A.~J. Sheldrick, U.~Hegerl, and H.~Himmerich}.
\newblock \href{https://doi.org/10.1186/1744-9081-8-50}{Time perception at
  different {EEG}-vigilance levels}.
\newblock \href{https://doi.org/10.1186/1744-9081-8-50}{{\em Behavioral and
  Brain Functions}}, \href{https://doi.org/10.1186/1744-9081-8-50}{8:1--8},
  \href{https://doi.org/10.1186/1744-9081-8-50}{Sept. 2012}.
  \href{https://doi.org/10.1186/1744-9081-8-50}
{doi: \textsf{%
10\hspace{.1pt}\discretionary{.}{%
}{.}\hspace{.4pt}1186\discretionary{/}{%
}{/}1744\discretionary{%
}{-}{-}9081\discretionary{%
}{-}{-}8\discretionary{%
}{-}{-}50}}


\bibitem{mioni2020modulation}
\href{https://doi.org/10.1093/texcom/tgaa064}{G.~Mioni, A.~Shelp, C.~T.
  Stanfield-Wiswell, K.~A. Gladhill, F.~Bader, and M.~Wiener}.
\newblock \href{https://doi.org/10.1093/texcom/tgaa064}{Modulation of
  individual alpha frequency with t{ACS} shifts time perception}.
\newblock \href{https://doi.org/10.1093/texcom/tgaa064}{{\em Cerebral Cortex
  Communications}},
  \href{https://doi.org/10.1093/texcom/tgaa064}{1(1):tgaa064},
  \href{https://doi.org/10.1093/texcom/tgaa064}{Sept. 2020}.
  \href{https://doi.org/10.1093/texcom/tgaa064}
{doi: \textsf{%
10\hspace{.1pt}\discretionary{.}{%
}{.}\hspace{.4pt}1093\discretionary{/}{%
}{/}texcom\discretionary{/}{%
}{/}tgaa064}}


\bibitem{moinnereau2023quantifying}
\href{https://doi.org/10.3389/fnrgo.2023.1189179}{M.-A. Moinnereau, A.~A.
  Oliveira, and T.~H. Falk}.
\newblock \href{https://doi.org/10.3389/fnrgo.2023.1189179}{Quantifying time
  perception during virtual reality gameplay using a multimodal
  biosensor-instrumented headset: a feasibility study}.
\newblock \href{https://doi.org/10.3389/fnrgo.2023.1189179}{{\em Frontiers in
  Neuroergonomics}},
  \href{https://doi.org/10.3389/fnrgo.2023.1189179}{4:1189179},
  \href{https://doi.org/10.3389/fnrgo.2023.1189179}{July 2023}.
  \href{https://doi.org/10.3389/fnrgo.2023.1189179}
{doi: \textsf{%
10\hspace{.1pt}\discretionary{.}{%
}{.}\hspace{.4pt}3389\discretionary{/}{%
}{/}fnrgo\hspace{.1pt}\discretionary{.}{%
}{.}\hspace{.4pt}2023\hspace{.1pt}\discretionary{.}{%
}{.}\hspace{.4pt}1189179}}


\bibitem{morin2024mindfulness}
\href{https://doi.org/10.1016/j.neubiorev.2024.105657}{A.~Morin and
  S.~Grondin}.
\newblock \href{https://doi.org/10.1016/j.neubiorev.2024.105657}{{M}indfulness
  and {T}ime {P}erception: A systematic integrative review}.
\newblock \href{https://doi.org/10.1016/j.neubiorev.2024.105657}{{\em
  Neuroscience \& Biobehavioral Reviews}},
  \href{https://doi.org/10.1016/j.neubiorev.2024.105657}{162:105657},
  \href{https://doi.org/10.1016/j.neubiorev.2024.105657}{July 2024}.
  \href{https://doi.org/10.1016/j.neubiorev.2024.105657}
{doi: \textsf{%
10\hspace{.1pt}\discretionary{.}{%
}{.}\hspace{.4pt}1016\discretionary{/}{%
}{/}j\hspace{.1pt}\discretionary{.}{%
}{.}\hspace{.4pt}neubiorev\hspace{.1pt}\discretionary{.}{%
}{.}\hspace{.4pt}2024\hspace{.1pt}\discretionary{.}{%
}{.}\hspace{.4pt}105657}}


\bibitem{mullen2021time}
\href{https://doi.org/10.1163/22134468-bja10034}{G.~Mullen and N.~Davidenko}.
\newblock \href{https://doi.org/10.1163/22134468-bja10034}{Time compression in
  virtual reality}.
\newblock \href{https://doi.org/10.1163/22134468-bja10034}{{\em Timing \& Time
  Perception}},
  \href{https://doi.org/10.1163/22134468-bja10034}{9(4):377--392},
  \href{https://doi.org/10.1163/22134468-bja10034}{May 2021}.
  \href{https://doi.org/10.1163/22134468-bja10034}
{doi: \textsf{%
10\hspace{.1pt}\discretionary{.}{%
}{.}\hspace{.4pt}1163\discretionary{/}{%
}{/}22134468\discretionary{%
}{-}{-}bja10034}}


\bibitem{nuyens2020potential}
\href{https://doi.org/10.1007/s11469-019-00121-1}{F.~M. Nuyens, D.~J. Kuss,
  O.~Lopez-Fernandez, and M.~D. Griffiths}.
\newblock \href{https://doi.org/10.1007/s11469-019-00121-1}{The potential
  interaction between time perception and gaming: A narrative review}.
\newblock \href{https://doi.org/10.1007/s11469-019-00121-1}{{\em International
  Journal of Mental Health and Addiction}},
  \href{https://doi.org/10.1007/s11469-019-00121-1}{18:1226--1246},
  \href{https://doi.org/10.1007/s11469-019-00121-1}{Aug. 2020}.
  \href{https://doi.org/10.1007/s11469-019-00121-1}
{doi: \textsf{%
10\hspace{.1pt}\discretionary{.}{%
}{.}\hspace{.4pt}1007\discretionary{/}{%
}{/}s11469\discretionary{%
}{-}{-}019\discretionary{%
}{-}{-}00121\discretionary{%
}{-}{-}1}}


\bibitem{pariyadath2007effect}
\href{https://doi.org/10.1371/journal.pone.0001264}{V.~Pariyadath and
  D.~Eagleman}.
\newblock \href{https://doi.org/10.1371/journal.pone.0001264}{The effect of
  predictability on subjective duration}.
\newblock \href{https://doi.org/10.1371/journal.pone.0001264}{{\em PLoS ONE}},
  \href{https://doi.org/10.1371/journal.pone.0001264}{2(11):e1264},
  \href{https://doi.org/10.1371/journal.pone.0001264}{Nov. 2007}.
  \href{https://doi.org/10.1371/journal.pone.0001264}
{doi: \textsf{%
10\hspace{.1pt}\discretionary{.}{%
}{.}\hspace{.4pt}1371\discretionary{/}{%
}{/}journal\hspace{.1pt}\discretionary{.}{%
}{.}\hspace{.4pt}pone\hspace{.1pt}\discretionary{.}{%
}{.}\hspace{.4pt}0001264}}


\bibitem{polti2018effect}
\href{https://doi.org/10.1038/s41598-018-25119-y}{I.~Polti, B.~Martin, and
  V.~van Wassenhove}.
\newblock \href{https://doi.org/10.1038/s41598-018-25119-y}{The effect of
  attention and working memory on the estimation of elapsed time}.
\newblock \href{https://doi.org/10.1038/s41598-018-25119-y}{{\em Scientific
  Reports}}, \href{https://doi.org/10.1038/s41598-018-25119-y}{8(1):6690},
  \href{https://doi.org/10.1038/s41598-018-25119-y}{Apr. 2018}.
  \href{https://doi.org/10.1038/s41598-018-25119-y}
{doi: \textsf{%
10\hspace{.1pt}\discretionary{.}{%
}{.}\hspace{.4pt}1038\discretionary{/}{%
}{/}s41598\discretionary{%
}{-}{-}018\discretionary{%
}{-}{-}25119\discretionary{%
}{-}{-}y}}


\bibitem{rau2006time}
\href{https://doi.org/10.1089/cpb.2006.9.396}{P.-L.~P. Rau, S.-Y. Peng, and
  C.-C. Yang}.
\newblock \href{https://doi.org/10.1089/cpb.2006.9.396}{Time distortion for
  expert and novice online game players}.
\newblock \href{https://doi.org/10.1089/cpb.2006.9.396}{{\em CyberPsychology \&
  Behavior}}, \href{https://doi.org/10.1089/cpb.2006.9.396}{9(4):396--403},
  \href{https://doi.org/10.1089/cpb.2006.9.396}{Aug. 2006}.
  \href{https://doi.org/10.1089/cpb.2006.9.396}
{doi: \textsf{%
10\hspace{.1pt}\discretionary{.}{%
}{.}\hspace{.4pt}1089\discretionary{/}{%
}{/}cpb\hspace{.1pt}\discretionary{.}{%
}{.}\hspace{.4pt}2006\hspace{.1pt}\discretionary{.}{%
}{.}\hspace{.4pt}9\hspace{.1pt}\discretionary{.}{%
}{.}\hspace{.4pt}396}}


\bibitem{read2021engagement}
\href{https://doi.org/10.1177/1071181321651337}{T.~Read, C.~A. Sanchez, and
  R.~De~Amicis}.
\newblock \href{https://doi.org/10.1177/1071181321651337}{Engagement and time
  perception in virtual reality}.
\newblock \href{https://doi.org/10.1177/1071181321651337}{In {\em Proc.\
  HFES}}, \href{https://doi.org/10.1177/1071181321651337}{vol.~65},
  \href{https://doi.org/10.1177/1071181321651337}{pp. 913--918}.
  \href{https://doi.org/10.1177/1071181321651337}{Sage Publications},
  \href{https://doi.org/10.1177/1071181321651337}{Los Angeles, CA, USA},
  \href{https://doi.org/10.1177/1071181321651337}{2021}.
  \href{https://doi.org/10.1177/1071181321651337}
{doi: \textsf{%
10\hspace{.1pt}\discretionary{.}{%
}{.}\hspace{.4pt}1177\discretionary{/}{%
}{/}1071181321651337}}


\bibitem{read2023influence}
\href{https://doi.org/10.1007/s10055-022-00723-6}{T.~Read, C.~A. Sanchez, and
  R.~De~Amicis}.
\newblock \href{https://doi.org/10.1007/s10055-022-00723-6}{The influence of
  attentional engagement and spatial characteristics on time perception in
  virtual reality}.
\newblock \href{https://doi.org/10.1007/s10055-022-00723-6}{{\em Virtual
  Reality}},
  \href{https://doi.org/10.1007/s10055-022-00723-6}{27(2):1265--1272},
  \href{https://doi.org/10.1007/s10055-022-00723-6}{Dec. 2023}.
  \href{https://doi.org/10.1007/s10055-022-00723-6}
{doi: \textsf{%
10\hspace{.1pt}\discretionary{.}{%
}{.}\hspace{.4pt}1007\discretionary{/}{%
}{/}s10055\discretionary{%
}{-}{-}022\discretionary{%
}{-}{-}00723\discretionary{%
}{-}{-}6}}


\bibitem{riegel2016characterization}
\href{https://doi.org/10.3758/s13428-015-0620-1}{M.~Riegel, {\L}.~{\.Z}urawski,
  M.~Wierzba, A.~Moslehi, {\L}.~Klocek, M.~Horvat, A.~Grabowska,
  J.~Micha{\l}owski, K.~Jednor{\'o}g, and A.~Marchewka}.
\newblock \href{https://doi.org/10.3758/s13428-015-0620-1}{Characterization of
  the {N}encki {A}ffective {P}icture {S}ystem by discrete emotional categories
  ({NAPS BE})}.
\newblock \href{https://doi.org/10.3758/s13428-015-0620-1}{{\em Behavior
  Research Methods}},
  \href{https://doi.org/10.3758/s13428-015-0620-1}{48:600--612},
  \href{https://doi.org/10.3758/s13428-015-0620-1}{July 2016}.
  \href{https://doi.org/10.3758/s13428-015-0620-1}
{doi: \textsf{%
10\hspace{.1pt}\discretionary{.}{%
}{.}\hspace{.4pt}3758\discretionary{/}{%
}{/}s13428\discretionary{%
}{-}{-}015\discretionary{%
}{-}{-}0620\discretionary{%
}{-}{-}1}}


\bibitem{schatzschneider2016turned}
\href{https://doi.org/10.1109/TVCG.2016.2518137}{C.~Schatzschneider, G.~Bruder,
  and F.~Steinicke}.
\newblock \href{https://doi.org/10.1109/TVCG.2016.2518137}{Who turned the
  clock? effects of manipulated zeitgebers, cognitive load and immersion on
  time estimation}.
\newblock \href{https://doi.org/10.1109/TVCG.2016.2518137}{{\em IEEE
  Transactions on Visualization and Computer Graphics}},
  \href{https://doi.org/10.1109/TVCG.2016.2518137}{22(4):1387--1395},
  \href{https://doi.org/10.1109/TVCG.2016.2518137}{Apr. 2016}.
  \href{https://doi.org/10.1109/TVCG.2016.2518137}
{doi: \textsf{%
10\hspace{.1pt}\discretionary{.}{%
}{.}\hspace{.4pt}1109\discretionary{/}{%
}{/}TVCG\hspace{.1pt}\discretionary{.}{%
}{.}\hspace{.4pt}2016\hspace{.1pt}\discretionary{.}{%
}{.}\hspace{.4pt}2518137}}


\bibitem{schneider2011effect}
\href{https://doi.org/10.1007/s00520-010-0852-7}{S.~M. Schneider, C.~K. Kisby,
  and E.~P. Flint}.
\newblock \href{https://doi.org/10.1007/s00520-010-0852-7}{Effect of virtual
  reality on time perception in patients receiving chemotherapy}.
\newblock \href{https://doi.org/10.1007/s00520-010-0852-7}{{\em Supportive Care
  in Cancer}},
  \href{https://doi.org/10.1007/s00520-010-0852-7}{19(4):555--564},
  \href{https://doi.org/10.1007/s00520-010-0852-7}{Mar. 2010}.
  \href{https://doi.org/10.1007/s00520-010-0852-7}
{doi: \textsf{%
10\hspace{.1pt}\discretionary{.}{%
}{.}\hspace{.4pt}1007\discretionary{/}{%
}{/}s00520\discretionary{%
}{-}{-}010\discretionary{%
}{-}{-}0852\discretionary{%
}{-}{-}7}}


\bibitem{schweitzer2017associated}
\href{https://doi.org/10.1177/0301006616689579}{R.~Schweitzer, S.~Trapp, and
  M.~Bar}.
\newblock \href{https://doi.org/10.1177/0301006616689579}{Associated
  information increases subjective perception of duration}.
\newblock \href{https://doi.org/10.1177/0301006616689579}{{\em Perception}},
  \href{https://doi.org/10.1177/0301006616689579}{46(8):1000--1007},
  \href{https://doi.org/10.1177/0301006616689579}{Jan. 2017}.
  \href{https://doi.org/10.1177/0301006616689579}
{doi: \textsf{%
10\hspace{.1pt}\discretionary{.}{%
}{.}\hspace{.4pt}1177\discretionary{/}{%
}{/}0301006616689579}}


\bibitem{simchy2018expectation}
\href{https://doi.org/10.3758/s13414-017-1432-4}{R.~Simchy-Gross and E.~H.
  Margulis}.
\newblock \href{https://doi.org/10.3758/s13414-017-1432-4}{Expectation,
  information processing, and subjective duration}.
\newblock \href{https://doi.org/10.3758/s13414-017-1432-4}{{\em Attention,
  Perception, \& Psychophysics}},
  \href{https://doi.org/10.3758/s13414-017-1432-4}{80:275--291},
  \href{https://doi.org/10.3758/s13414-017-1432-4}{Oct. 2018}.
  \href{https://doi.org/10.3758/s13414-017-1432-4}
{doi: \textsf{%
10\hspace{.1pt}\discretionary{.}{%
}{.}\hspace{.4pt}3758\discretionary{/}{%
}{/}s13414\discretionary{%
}{-}{-}017\discretionary{%
}{-}{-}1432\discretionary{%
}{-}{-}4}}


\bibitem{slater1993representation}
\href{https://doi.org/10.1162/pres.1993.2.3.221}{M.~Slater and M.~Usoh}.
\newblock \href{https://doi.org/10.1162/pres.1993.2.3.221}{Representation
  systems, perceptual position and presence in virtual environments}.
\newblock \href{https://doi.org/10.1162/pres.1993.2.3.221}{{\em Presence:
  Teleoperators and Virtual Environments}},
  \href{https://doi.org/10.1162/pres.1993.2.3.221}{2(3):221--234},
  \href{https://doi.org/10.1162/pres.1993.2.3.221}{Aug. 1993}.
  \href{https://doi.org/10.1162/pres.1993.2.3.221}
{doi: \textsf{%
10\hspace{.1pt}\discretionary{.}{%
}{.}\hspace{.4pt}1162\discretionary{/}{%
}{/}pres\hspace{.1pt}\discretionary{.}{%
}{.}\hspace{.4pt}1993\hspace{.1pt}\discretionary{.}{%
}{.}\hspace{.4pt}2\hspace{.1pt}\discretionary{.}{%
}{.}\hspace{.4pt}3\hspace{.1pt}\discretionary{.}{%
}{.}\hspace{.4pt}221}}


\bibitem{smith2011effects}
\href{https://doi.org/10.1037/a0026145}{S.~D. Smith, T.~A. McIver, M.~S.~J.
  Di~Nella, and M.~L. Crease}.
\newblock \href{https://doi.org/10.1037/a0026145}{The effects of valence and
  arousal on the emotional modulation of time perception: Evidence for multiple
  stages of processing}.
\newblock \href{https://doi.org/10.1037/a0026145}{{\em Emotion}},
  \href{https://doi.org/10.1037/a0026145}{11(6):1305},
  \href{https://doi.org/10.1037/a0026145}{Dec. 2011}.
  \href{https://doi.org/10.1037/a0026145}
{doi: \textsf{%
10\hspace{.1pt}\discretionary{.}{%
}{.}\hspace{.4pt}1037\discretionary{/}{%
}{/}a0026145}}


\bibitem{soderstrom2018users}
\href{https://doi.org/10.1145/3232078.3232092}{U.~S{\"o}derstr{\"o}m,
  M.~B{\aa}{\aa}th, and T.~Mejtoft}.
\newblock \href{https://doi.org/10.1145/3232078.3232092}{The users' time
  perception: The effect of various animation speeds on loading screens}.
\newblock \href{https://doi.org/10.1145/3232078.3232092}{In {\em Proc.\ ECCE}},
  \href{https://doi.org/10.1145/3232078.3232092}{pp. 1--4}.
  \href{https://doi.org/10.1145/3232078.3232092}{ACM},
  \href{https://doi.org/10.1145/3232078.3232092}{New York, NY, USA},
  \href{https://doi.org/10.1145/3232078.3232092}{2018}.
  \href{https://doi.org/10.1145/3232078.3232092}
{doi: \textsf{%
10\hspace{.1pt}\discretionary{.}{%
}{.}\hspace{.4pt}1145\discretionary{/}{%
}{/}3232078\hspace{.1pt}\discretionary{.}{%
}{.}\hspace{.4pt}3232092}}


\bibitem{sunthorn2023making}
\href{https://doi.org/10.1080/0144929X.2022.2161937}{W.~Sunthorn,
  C.~Le~Mercier, and C.~Silpasuwanchai}.
\newblock \href{https://doi.org/10.1080/0144929X.2022.2161937}{Making time
  perception shorter with pitch and interval patterns}.
\newblock \href{https://doi.org/10.1080/0144929X.2022.2161937}{{\em Behaviour
  \& Information Technology}},
  \href{https://doi.org/10.1080/0144929X.2022.2161937}{43(2):273--283},
  \href{https://doi.org/10.1080/0144929X.2022.2161937}{Jan. 2023}.
  \href{https://doi.org/10.1080/0144929X.2022.2161937}
{doi: \textsf{%
10\hspace{.1pt}\discretionary{.}{%
}{.}\hspace{.4pt}1080\discretionary{/}{%
}{/}0144929X\hspace{.1pt}\discretionary{.}{%
}{.}\hspace{.4pt}2022\hspace{.1pt}\discretionary{.}{%
}{.}\hspace{.4pt}2161937}}


\bibitem{tachmatzidou2023attention}
\href{https://doi.org/10.1038/s41598-023-37030-2}{O.~Tachmatzidou and
  A.~Vatakis}.
\newblock \href{https://doi.org/10.1038/s41598-023-37030-2}{Attention and
  schema violations of real world scenes differentially modulate time
  perception}.
\newblock \href{https://doi.org/10.1038/s41598-023-37030-2}{{\em Scientific
  Reports}}, \href{https://doi.org/10.1038/s41598-023-37030-2}{13(1):10002},
  \href{https://doi.org/10.1038/s41598-023-37030-2}{June 2023}.
  \href{https://doi.org/10.1038/s41598-023-37030-2}
{doi: \textsf{%
10\hspace{.1pt}\discretionary{.}{%
}{.}\hspace{.4pt}1038\discretionary{/}{%
}{/}s41598\discretionary{%
}{-}{-}023\discretionary{%
}{-}{-}37030\discretionary{%
}{-}{-}2}}


\bibitem{triberti2018matters}
\href{https://doi.org/10.1016/j.abrep.2018.06.003}{S.~Triberti, L.~Milani,
  D.~Villani, S.~Grumi, S.~Peracchia, G.~Curcio, and G.~Riva}.
\newblock \href{https://doi.org/10.1016/j.abrep.2018.06.003}{What matters is
  when you play: Investigating the relationship between online video games
  addiction and time spent playing over specific day phases}.
\newblock \href{https://doi.org/10.1016/j.abrep.2018.06.003}{{\em Addictive
  Behaviors Reports}},
  \href{https://doi.org/10.1016/j.abrep.2018.06.003}{8:185--188},
  \href{https://doi.org/10.1016/j.abrep.2018.06.003}{Dec. 2018}.
  \href{https://doi.org/10.1016/j.abrep.2018.06.003}
{doi: \textsf{%
10\hspace{.1pt}\discretionary{.}{%
}{.}\hspace{.4pt}1016\discretionary{/}{%
}{/}j\hspace{.1pt}\discretionary{.}{%
}{.}\hspace{.4pt}abrep\hspace{.1pt}\discretionary{.}{%
}{.}\hspace{.4pt}2018\hspace{.1pt}\discretionary{.}{%
}{.}\hspace{.4pt}06\hspace{.1pt}\discretionary{.}{%
}{.}\hspace{.4pt}003}}


\bibitem{tse2004attention}
\href{https://doi.org/10.3758/bf03196844}{P.~U. Tse, J.~Intriligator,
  J.~Rivest, and P.~Cavanagh}.
\newblock \href{https://doi.org/10.3758/bf03196844}{Attention and the
  subjective expansion of time}.
\newblock \href{https://doi.org/10.3758/bf03196844}{{\em Perception \&
  Psychophysics}}, \href{https://doi.org/10.3758/bf03196844}{66(7):1171--1189},
  \href{https://doi.org/10.3758/bf03196844}{Oct. 2004}.
  \href{https://doi.org/10.3758/bf03196844}
{doi: \textsf{%
10\hspace{.1pt}\discretionary{.}{%
}{.}\hspace{.4pt}3758\discretionary{/}{%
}{/}bf03196844}}


\bibitem{ulrich2006perceived}
\href{https://doi.org/10.1007/s00426-004-0195-4}{R.~Ulrich, J.~Nitschke, and
  T.~Rammsayer}.
\newblock \href{https://doi.org/10.1007/s00426-004-0195-4}{Perceived duration
  of expected and unexpected stimuli}.
\newblock \href{https://doi.org/10.1007/s00426-004-0195-4}{{\em Psychological
  research}}, \href{https://doi.org/10.1007/s00426-004-0195-4}{70:77--87},
  \href{https://doi.org/10.1007/s00426-004-0195-4}{Dec. 2006}.
  \href{https://doi.org/10.1007/s00426-004-0195-4}
{doi: \textsf{%
10\hspace{.1pt}\discretionary{.}{%
}{.}\hspace{.4pt}1007\discretionary{/}{%
}{/}s00426\discretionary{%
}{-}{-}004\discretionary{%
}{-}{-}0195\discretionary{%
}{-}{-}4}}


\bibitem{unruh2021influence}
\href{https://doi.org/10.3389/frvir.2021.658509}{F.~Unruh, M.~Landeck,
  S.~Oberd{\"o}rfer, J.-L. Lugrin, and M.~E. Latoschik}.
\newblock \href{https://doi.org/10.3389/frvir.2021.658509}{The influence of
  avatar embodiment on time perception - towards {VR} for time-based therapy}.
\newblock \href{https://doi.org/10.3389/frvir.2021.658509}{{\em Frontiers in
  Virtual Reality}},
  \href{https://doi.org/10.3389/frvir.2021.658509}{2:658509},
  \href{https://doi.org/10.3389/frvir.2021.658509}{July 2021}.
  \href{https://doi.org/10.3389/frvir.2021.658509}
{doi: \textsf{%
10\hspace{.1pt}\discretionary{.}{%
}{.}\hspace{.4pt}3389\discretionary{/}{%
}{/}frvir\hspace{.1pt}\discretionary{.}{%
}{.}\hspace{.4pt}2021\hspace{.1pt}\discretionary{.}{%
}{.}\hspace{.4pt}658509}}


\bibitem{unruh2023body}
\href{https://doi.org/10.1109/TVCG.2023.3247040}{F.~Unruh, D.~Vogel,
  M.~Landeck, J.-L. Lugrin, and M.~E. Latoschik}.
\newblock \href{https://doi.org/10.1109/TVCG.2023.3247040}{{B}ody and {T}ime:
  Virtual embodiment and its effect on time perception}.
\newblock \href{https://doi.org/10.1109/TVCG.2023.3247040}{{\em IEEE
  Transactions on Visualization and Computer Graphics}},
  \href{https://doi.org/10.1109/TVCG.2023.3247040}{29(5):2626--2636},
  \href{https://doi.org/10.1109/TVCG.2023.3247040}{May 2023}.
  \href{https://doi.org/10.1109/TVCG.2023.3247040}
{doi: \textsf{%
10\hspace{.1pt}\discretionary{.}{%
}{.}\hspace{.4pt}1109\discretionary{/}{%
}{/}TVCG\hspace{.1pt}\discretionary{.}{%
}{.}\hspace{.4pt}2023\hspace{.1pt}\discretionary{.}{%
}{.}\hspace{.4pt}3247040}}


\bibitem{utegaliyev2024expected}
\href{https://doi.org/10.1037/xhp0001179}{N.~Utegaliyev and C.~von Castell}.
\newblock \href{https://doi.org/10.1037/xhp0001179}{Expected events dilate
  subjective duration in the auditory modality: Effects of predictability and
  expectation on time perception}.
\newblock \href{https://doi.org/10.1037/xhp0001179}{{\em Journal of
  Experimental Psychology: Human Perception and Performance}},
  \href{https://doi.org/10.1037/xhp0001179}{50(3):249},
  \href{https://doi.org/10.1037/xhp0001179}{Mar. 2024}.
  \href{https://doi.org/10.1037/xhp0001179}
{doi: \textsf{%
10\hspace{.1pt}\discretionary{.}{%
}{.}\hspace{.4pt}1037\discretionary{/}{%
}{/}xhp0001179}}


\bibitem{wiener2018intrinsic}
\href{https://doi.org/10.1038/s41598-018-26385-6}{M.~Wiener, A.~Parikh,
  A.~Krakow, and H.~B. Coslett}.
\newblock \href{https://doi.org/10.1038/s41598-018-26385-6}{An intrinsic role
  of beta oscillations in memory for time estimation}.
\newblock \href{https://doi.org/10.1038/s41598-018-26385-6}{{\em Scientific
  Reports}}, \href{https://doi.org/10.1038/s41598-018-26385-6}{8(1):7992},
  \href{https://doi.org/10.1038/s41598-018-26385-6}{May 2018}.
  \href{https://doi.org/10.1038/s41598-018-26385-6}
{doi: \textsf{%
10\hspace{.1pt}\discretionary{.}{%
}{.}\hspace{.4pt}1038\discretionary{/}{%
}{/}s41598\discretionary{%
}{-}{-}018\discretionary{%
}{-}{-}26385\discretionary{%
}{-}{-}6}}


\bibitem{wood2007time}
\href{https://doi.org/10.1007/s11469-006-9048-2}{R.~T.~A. Wood and M.~D.
  Griffiths}.
\newblock \href{https://doi.org/10.1007/s11469-006-9048-2}{Time loss whilst
  playing video games: Is there a relationship to addictive behaviours?}
\newblock \href{https://doi.org/10.1007/s11469-006-9048-2}{{\em International
  Journal of Mental Health and Addiction}},
  \href{https://doi.org/10.1007/s11469-006-9048-2}{5:141--149},
  \href{https://doi.org/10.1007/s11469-006-9048-2}{Jan. 2007}.
  \href{https://doi.org/10.1007/s11469-006-9048-2}
{doi: \textsf{%
10\hspace{.1pt}\discretionary{.}{%
}{.}\hspace{.4pt}1007\discretionary{/}{%
}{/}s11469\discretionary{%
}{-}{-}006\discretionary{%
}{-}{-}9048\discretionary{%
}{-}{-}2}}


\bibitem{xiang2021confidence}
\href{https://doi.org/10.3758/s13414-021-02300-6}{Y.~Xiang, T.~Graeber,
  B.~Enke, and S.~J. Gershman}.
\newblock \href{https://doi.org/10.3758/s13414-021-02300-6}{Confidence and
  central tendency in perceptual judgment}.
\newblock \href{https://doi.org/10.3758/s13414-021-02300-6}{{\em Attention,
  Perception, \& Psychophysics}},
  \href{https://doi.org/10.3758/s13414-021-02300-6}{83:3024--3034},
  \href{https://doi.org/10.3758/s13414-021-02300-6}{Apr. 2021}.
  \href{https://doi.org/10.3758/s13414-021-02300-6}
{doi: \textsf{%
10\hspace{.1pt}\discretionary{.}{%
}{.}\hspace{.4pt}3758\discretionary{/}{%
}{/}s13414\discretionary{%
}{-}{-}021\discretionary{%
}{-}{-}02300\discretionary{%
}{-}{-}6}}


\bibitem{xuan2007larger}
\href{https://doi.org/10.1167/7.10.2}{B.~Xuan, D.~Zhang, S.~He, and X.~Chen}.
\newblock \href{https://doi.org/10.1167/7.10.2}{Larger stimuli are judged to
  last longer}.
\newblock \href{https://doi.org/10.1167/7.10.2}{{\em Journal of Vision}},
  \href{https://doi.org/10.1167/7.10.2}{7(10):2--2},
  \href{https://doi.org/10.1167/7.10.2}{July 2007}.
  \href{https://doi.org/10.1167/7.10.2}
{doi: \textsf{%
10\hspace{.1pt}\discretionary{.}{%
}{.}\hspace{.4pt}1167\discretionary{/}{%
}{/}7\hspace{.1pt}\discretionary{.}{%
}{.}\hspace{.4pt}10\hspace{.1pt}\discretionary{.}{%
}{.}\hspace{.4pt}2}}


\bibitem{yang2020high}
\href{https://doi.org/10.3389/fnhum.2020.00089}{K.~Yang, L.~Tong, J.~Shu,
  N.~Zhuang, B.~Yan, and Y.~Zeng}.
\newblock \href{https://doi.org/10.3389/fnhum.2020.00089}{High {G}amma {B}and
  {EEG} {C}losely {R}elated to {E}motion: {E}vidence from functional network}.
\newblock \href{https://doi.org/10.3389/fnhum.2020.00089}{{\em Frontiers in
  human neuroscience}}, \href{https://doi.org/10.3389/fnhum.2020.00089}{14:89},
  \href{https://doi.org/10.3389/fnhum.2020.00089}{Mar. 2020}.
  \href{https://doi.org/10.3389/fnhum.2020.00089}
{doi: \textsf{%
10\hspace{.1pt}\discretionary{.}{%
}{.}\hspace{.4pt}3389\discretionary{/}{%
}{/}fnhum\hspace{.1pt}\discretionary{.}{%
}{.}\hspace{.4pt}2020\hspace{.1pt}\discretionary{.}{%
}{.}\hspace{.4pt}00089}}


\bibitem{zakay1989subjective}
\href{https://doi.org/10.1016/s0166-4115(08)61047-x}{D.~Zakay}.
\newblock \href{https://doi.org/10.1016/s0166-4115(08)61047-x}{Subjective
  {T}ime and {A}ttentional {R}esource {A}llocation: {A}n integrated model of
  time estimation}.
\newblock \href{https://doi.org/10.1016/s0166-4115(08)61047-x}{In I.~Levin and
  D.~Zakay, eds., {\em Time and Human Cognition: A Life-Span Perspective}},
  \href{https://doi.org/10.1016/s0166-4115(08)61047-x}{pp. 365--397}.
  \href{https://doi.org/10.1016/s0166-4115(08)61047-x}{Elsevier},
  \href{https://doi.org/10.1016/s0166-4115(08)61047-x}{Amsterdam, The
  Netherlands}, \href{https://doi.org/10.1016/s0166-4115(08)61047-x}{1989}.
  \href{https://doi.org/10.1016/s0166-4115(08)61047-x}
{doi: \textsf{%
10\hspace{.1pt}\discretionary{.}{%
}{.}\hspace{.4pt}1016\discretionary{/}{%
}{/}s0166\discretionary{%
}{-}{-}4115\discretionary{%
}{(}{(}08\discretionary{)}{%
}{)}61047\discretionary{%
}{-}{-}x}}


\bibitem{zakay1993relative}
\href{https://doi.org/10.3758/bf03211789}{D.~Zakay}.
\newblock \href{https://doi.org/10.3758/bf03211789}{Relative and absolute
  duration judgments under prospective and retrospective paradigms}.
\newblock \href{https://doi.org/10.3758/bf03211789}{{\em Perception \&
  Psychophysics}}, \href{https://doi.org/10.3758/bf03211789}{54(5):656--664},
  \href{https://doi.org/10.3758/bf03211789}{Sept. 1993}.
  \href{https://doi.org/10.3758/bf03211789}
{doi: \textsf{%
10\hspace{.1pt}\discretionary{.}{%
}{.}\hspace{.4pt}3758\discretionary{/}{%
}{/}bf03211789}}


\bibitem{zakay2014psychological}
\href{https://doi.org/10.3389/fpsyg.2014.00917}{D.~Zakay}.
\newblock \href{https://doi.org/10.3389/fpsyg.2014.00917}{Psychological time as
  information: The case of boredom}.
\newblock \href{https://doi.org/10.3389/fpsyg.2014.00917}{{\em Frontiers in
  Psychology}}, \href{https://doi.org/10.3389/fpsyg.2014.00917}{5:917},
  \href{https://doi.org/10.3389/fpsyg.2014.00917}{Aug. 2014}.
  \href{https://doi.org/10.3389/fpsyg.2014.00917}
{doi: \textsf{%
10\hspace{.1pt}\discretionary{.}{%
}{.}\hspace{.4pt}3389\discretionary{/}{%
}{/}fpsyg\hspace{.1pt}\discretionary{.}{%
}{.}\hspace{.4pt}2014\hspace{.1pt}\discretionary{.}{%
}{.}\hspace{.4pt}00917}}


\end{thebibliography}

\end{document}